\documentclass[aps,tightenlines,twocolumn,floats,prd,nofootinbib,
superscriptaddress,showpacs,preprintnumbers]{revtex4-1}
\def\bea{\begin{eqnarray}}
\def\be{\begin{equation}}
\def\ee{\end{equation}}
\def\eea{\end{eqnarray}}
\def\bal{\begin{align}}
\def\eal{\end{align}}

\def\sfrac#1#2{{\textstyle \frac{#1}{#2}}}

\usepackage[usenames,dvipsnames]{color}
\usepackage{graphics}
\usepackage{graphicx}
\usepackage{epsf} 
\usepackage{amsmath}
\usepackage{amssymb}
\usepackage{bbold}
\usepackage{bm}
\usepackage{slashed}
\usepackage{float}
\usepackage{footnote} 
\usepackage{xcolor}
\usepackage{textcomp}
\def\peq #1{\scriptstyle{#1}}

\begin{document}
 \preprint{CFTP/15-009}
 \preprint{JLAB-THY-15-2125}

\title{Charge-conjugation symmetric complete impulse approximation for the \\pion electromagnetic form factor in the Covariant Spectator Theory}



\author{Elmar P. Biernat}
\email[]{elmar.biernat@tecnico.ulisboa.pt}
\affiliation{Centro de F\'isica Te\'orica de Part\'iculas (CFTP), 
Instituto Superior T\'ecnico (IST), Universidade de Lisboa, 
Av. Rovisco Pais, 1049-001 Lisboa, Portugal}

\author{Franz Gross }
\email[]{gross@jlab.org}
 \affiliation{ Thomas Jefferson National Accelerator Facility (JLab), 
Newport News, VA 23606, USA}
\affiliation{College of William and Mary, Williamsburg, VA 23188, USA}

\author{ M. T. Pe\~na}
\email[]{teresa.pena@tecnico.ulisboa.pt}
\affiliation{Centro de F\'isica Te\'orica de Part\'iculas (CFTP), 
Instituto Superior T\'ecnico (IST), Universidade de Lisboa, 
Av. Rovisco Pais, 1049-001 Lisboa, Portugal}

\author{Alfred Stadler}
\email[]{stadler@uevora.pt}
\affiliation{Departamento de F\'isica, Universidade de 
\'Evora, 7000-671 \'Evora, Portugal} 
\affiliation{Centro de F\'isica Te\'orica de Part\'iculas (CFTP), 
Instituto Superior T\'ecnico (IST), Universidade de Lisboa, 
Av. Rovisco Pais, 1049-001 Lisboa, Portugal}

\date{\today}

\begin{abstract}
The pion form factor  is calculated in the framework of the charge-conjugation invariant Covariant Spectator Theory. This formalism is established in Minkowski space and the calculation is set up in momentum space. In a previous calculation we included only the leading pole coming from the spectator quark (referred to as the relativistic impulse approximation). In this paper we also include the contributions from the poles of the quark which interacts with the photon and average over all poles in both the upper and lower half planes in order to preserve charge conjugation invariance (referred to as the $C$-symmetric complete impulse approximation). We find that for small pion mass these contributions are significant at all values of the four-momentum transfer $Q^2$ but, surprisingly, do not alter the shape obtained from the spectator poles alone.

\end{abstract}


\maketitle

\section{\label{sec:intro} Introduction }

With the 12 GeV accelerator upgrade at Jefferson Lab, the charged pion form factor $F_{\pi}$ will be known with high precision up to momentum transfer $Q^2 \approx 6$ GeV$^2$~\cite{Dudek:2012fk}. This measurement  will cover the interesting region where the product $Q^2 F_{\pi}$, as a function of $Q^2$, reaches a maximum and afterwards flattens down, and will help resolve the current discrepancy between the results for the $\pi \gamma^* \gamma$ transition form factor obtained by the Babar and Belle Collaborations.

In combination with theoretical calculations, the new measurements will narrow the uncertainty about the smallest momentum transfer at which the description based on asymptotic parton distribution functions is still valid. One may also certainly expect that the forthcoming data will clarify the mismatch between physical reality and perturbative QCD predictions in the region around $Q^2 \approx 6$ GeV$^2$~\cite{Chang:2013qy}. This paper focuses on the pion form factor in the small $Q^2$ region that is planned to be covered by the new experiment.  Part of the interest on this region lies also in its vicinity to the timelike sector and on the extrapolation of the derivative of the form factor to $Q^2 <0$.
The knowledge of the pion form factor enters into the evaluation of baryon form factors in the region near $Q^2 \approx 0$, and its behavior in the timelike region helps the interpretation of dilepton production data from heavy ion collisions.

Theoretically, the pion is the consequence both of the non-perturbative character of a quark-antiquark bound state, and of
Spontaneous Chiral-Symmetry Breaking (S$\chi$SB). Its role in nuclear structure and nuclear dynamics is crucial. The pion cloud is seen to contribute to the structure of the nucleon and its excitations, through the coupling to external photons. Also, the exchange of pions between nucleons dominates their interaction at large distances and is the primary origin of the  tensor force that has a decisive influence on the structure of nuclei.

The non-perturbative dynamics of the pion and other hadronic systems has been addressed by constituent quark models~\cite{Godfrey,Eichten:1975,Eichten:1978,Richardson:1978bt} and  QCD sum rules. These approaches do not provide a comprehensive and global view for both light  and heavy mesons, and baryons, and they cannot avoid a delicate fine-tuning between many parameters. More recently, QCD simulations on the lattice~\cite{Edwards,Guo}, light-front formulations of quantum field theory~\cite{Brodsky:1997de,Carbonell:1998rj,Sales:1999ec}, as well as models based on the Dyson-Schwinger approach and 
mass gap equation~\cite{Bars:1977ud,Amer:1983qa,LeYaouanc:1983it,Bicudo:1989sj,Alkofer:2000wg,Maris:2003vk,Fischer:2006ub,Rojas:2013tza}, start to provide an integrated account of mesons and baryons. 

In this paper we use the Covariant Spectator Theory (CST). In common with the Dyson-Schwinger framework, it generates a dynamical quark mass, which is a  function of the momentum, and this dressed mass is consistent with the two-body quark-antiquark dynamics. Although the CST equations share with lattice QCD and Dyson-Schwinger equations this important dynamical consistency, in contrast to those approaches CST equations are solved in Minkowski space. Therefore in this formalism the extension of results from the spacelike to the timelike region does not imply further work or the use of a different representation.

In Ref.~\cite{PhysRevD.89.016006} we presented the first calculation of the pion form factor based on the CST Bethe-Salpeter equation (CST-BSE) and the CST Dyson equation (CST-DE), using a  dressed quark mass function calibrated to fit existing lattice QCD data \cite{PhysRevD.89.016005}. 
In the present paper, as well as in Ref.~\cite{PhysRevD.89.016006}, the CST interaction kernel in momentum space has the form of a $\delta$-function {\it plus} a covariant generalization of the linear confining interaction. The fact that this model satisfies chiral symmetry is worth emphasizing; this is ensured by choosing a relativistic generalization of the confining  interaction ${\cal V}_L$ which decouples, in the chiral limit, from  both the one-body CST-DE and from  the two-body CST-BSE. In the calculation of the pion form factor in Ref.~\cite{PhysRevD.89.016006}  we used the relativistic impulse approximation (RIA). However, we know this approximation will 
break down for small pion masses, since under these conditions the pole contributions of the struck quark, which are omitted in RIA, are not negligible \cite{PhysRevD.89.016006}. The contribution of these additional poles to the charged pion form factor are calculated here, completing the results obtained from the RIA. In addition, we average over all of the propagator poles in both the upper and lower half planes, making our calculation consistent with the charge-conjugation symmetric equations from which the pion vertex must be calculated \cite{PhysRevD.89.016005}. This improvement is referred to as the $C$-symmetric omplete impulse approximation (C-CIA). A still more exact result can be obtained by adding interaction currents, a dynamically calculated pion vertex function, and the full dressing of the quark current to the C-CIA.  These are planned for future work (for more discussion, see the final section).

This paper is organized as follows: In Section \ref{sec:intro1} a brief review of the CST formalism is given. In Section~\ref{sec:triangleandingredients} we present the ingredients for the calculation of the triangle diagram, and in Section~\ref{sec:quark current} the formulas for the contributions of all the poles to the pion form factor. In Section \ref{sec:results} we present the results and compare to the ones obtained with RIA. Finally, Section \ref{sec:summary} presents a short summary and conclusions.

\section{\label{sec:intro1} Brief review of the one- and two-body CST equations}

In this section we briefly describe how the
CST is applied to the description of quark-antiquark mesons.
For technical details,  and references to earlier work, see ~\cite{Gross:1991te,Gross:1991pk,Gross:1994he,Savkli:1999me,PhysRevD.89.016005,PhysRevD.89.016006}.

In the four-dimensional BSE~\cite{Sal51} for 
heavy-light mesons, it is known~\cite{Gross:1993zj} that cancellations 
occur between  iterations of ladder diagrams and higher-order crossed-ladder diagrams in the complete kernel. The omission of crossed-ladder diagrams and part of the 
pole contributions of the ladder diagrams from the kernel can therefore give a better approximation 
to the exact BSE than the ladder approximation does. Also, efficiency in the summation of the series can be gained by recognizing these partial cancellations already at the outset. The fundamental idea of CST is then to reorganize the Bethe-Salpeter series into an equivalent form---the CST equation. This leads to a redefinition of  \textit{both} the complete kernel and the (off-mass-shell) two-particle propagators in the intermediate states. The resulting three-dimensional equation, the one-channel CST (or Gross)  Bethe-Salpeter equation~\cite{Gro69},  CST-BSE for short, is manifestly covariant. An important feature is that---unlike the BSE in ladder approximation---the CST-BSE has a smooth nonrelativistic limit, defining a natural covariant extension of the quantum mechanical Dirac and Schr\"{o}dinger equations to quantum field theory. 

In the heavy-light mass case, the redefined propagators are obtained by keeping only the positive-energy pole contribution from the heavy quark propagator in the energy loop integration. The heavy quark is then on its  positive-energy mass shell.  

In the case of two light quarks, the CST-BSE includes an explicit charge-conjugation symmetrization. The vertex functions of $\pi^+$ and $\pi^-$ are connected by charge conjugation and, therefore, both positive- and negative-energy quark poles must be included. This is realized by effectively averaging over the sums of all quark poles in the lower and in the upper complex-energy half plane, which generates the charge-conjugation-symmetric CST-BSE \cite{Savkli:1999me,PhysRevD.89.016005}, \vspace{0.1in}
 \begin{widetext} 
\bea
\Gamma(p_1,p_2)=-&&\frac12Z_0 \int_k
\Big[{\cal V}(\rho,\hat k- P/2)\Lambda(\hat k)\Gamma(\hat k,
\hat k-  P)S(\hat k-  P)
+{\cal V}(\rho,\hat k+P/2) S(\hat k+  P)\Gamma(\hat k+  P,  
\hat k)\Lambda(\hat k)
\nonumber\\&&
+{\cal V}(\rho,-\hat k-P/2) \Lambda(-\hat k)\Gamma(-\hat k,  -
\hat k-P)S(-\hat k-  P)
+{\cal V}(\rho,-\hat k+P/2) S(-\hat k+  P)\Gamma(-\hat k+  P,  -
\hat k)\Lambda(-\hat k)\Big]\,
\nonumber\\\equiv&& \,\mathrm i\int_{k0}
{\cal V}(\rho,\kappa) 
S(\kappa +P/2)\,\Gamma(\kappa+P/2,\kappa-P/2)\,S(\kappa-P/2)\,, 
\label{eq:CST-BSE}
\eea
\end{widetext}
where the three-dimensional covariant integration volume element is
\bea
\int_{\peq {k}} \equiv \int  \frac{\mathrm d^3 k}{(2\pi)^3} \frac{m}{E_k}\,,
\eea
and
\bea
\mathrm i\int_{k0} \equiv \mathrm i\int\frac{\mathrm d^4k}{(2\pi)^4}\,\bigg|_{\footnotesize\begin{array}{l}k_0\;\text{propagator} 
\cr \text{poles only}\end{array}}=-\frac12\sum_{\footnotesize\begin{array}{c}\text{propagator}
\cr \text{pole terms}\end{array}}\int_{\peq k} \,.\nonumber\\\label{eq:pp}
\eea
In Eq.~(\ref{eq:CST-BSE}) $\Gamma (p_1,p_2)$  is the ($4\times 4$) bound-state 
vertex function
with $p_1=\rho+\sfrac P2$  and $-p_2=-\rho+\sfrac P2$ the four-momenta of the outgoing quark and antiquark (respectively); $k_1=\kappa+\sfrac P2$ and $-k_2=-\kappa+\sfrac P2$ are the intermediate four-momenta for the quark and antiquark (respectively);  $P$ is the total bound-state momentum, $\rho$ and $\kappa$ are the relative momenta; $\hat k=(E_k,{\bf k})$ is the on-shell four-momentum with $E_k=\sqrt{m^2+{\bf k}^2}$; $\mathcal V (\rho,\kappa)\equiv \mathcal V (\rho,\kappa;P)$ is the interaction 
kernel with the general structure
\bea
 {\cal V}(\rho,\kappa) {\cal X}\equiv\sum_i V_i(\rho,\kappa){\cal O}_i{\cal X} {\cal O}_i \,,\label{eq:Vdecomp}
 \eea
 where the sum $i=\{ S, P, V, A, T\}$ is over the five possible invariant structures that could contribute: scalar, pseudoscalar, vector, axial-vector, and tensor.  The dressed quark  propagator, $S(k)$, projection operator, $\Lambda(k)$, and the numerator function, $N(k)$, are given by
 \bea 
S(k)&=&\frac1{m_{0}-\slashed{k}+\Sigma(k)-\mathrm i\epsilon}\,,\label{eq:dressedprop}\\
\Lambda(k)&=&\frac{M(k^2)+\slashed k}{2 M(k^2)} = \frac{ N(k)}{2 M(k^2)}
\, ,
\eea
where $M(k^2)$ the dressed quark mass function, $m_{0}$ the bare quark mass, and $\Sigma (k)$ is the quark self-energy.  For consistency, $\Sigma (k)$ is the solution of the one-body CST-DE using  the \textit{same} interaction kernel $\mathcal V$ that 
dresses the quark-antiquark vertex. One obtains~\cite{PhysRevD.89.016005} 
\bea
\Sigma (p)&=& \frac12Z_0\int_k
\Big\{{\cal V}(p,\hat k)\Lambda(\hat k)+{\cal V}(p,-\hat k)\Lambda(-\hat k)\Big\}
\nonumber\\&\equiv& -\mathrm i\int_{k0}
{\cal V}(p,k) S(k)\,.\label{eq:CST-DE}
\eea
The self-energy is written
\bea
\Sigma(p)=A(p^2)+\slashed{p} B(p^2)
\eea
and then 
\bea\label{eq:dressedprop1}
S(p)= Z(p^2)\frac{M(p^2)+\slashed{p}}{M^2 (p^2)-p^2-\mathrm i\epsilon} \equiv Z(p^2)\frac{N(p)}{D(p^2)}, 
\eea
where $D(p^2)=M^2 (p^2)-p^2-\mathrm i\epsilon$ is the denominator of the propagator, and the mass function $M(p^2)$ and the wave function normalization $Z(p^2)$ are
\bea
M(p^2)&=&\frac{A(p^2)+m_0}{1-B(p^2)}\,,
\nonumber\\
Z(p^2)&=&\frac{1}{1-B(p^2)}\,,
\eea
and $Z_0\equiv Z(m^2)$. For $\Sigma (p)=0$, $S(p)$ becomes the bare propagator denoted as $S_0(p)$.  For the models we are using, $B(p^2)=0$ and $Z(p^2)=1$.

\begin{figure}
\begin{center}
    \includegraphics[width=0.45\textwidth]{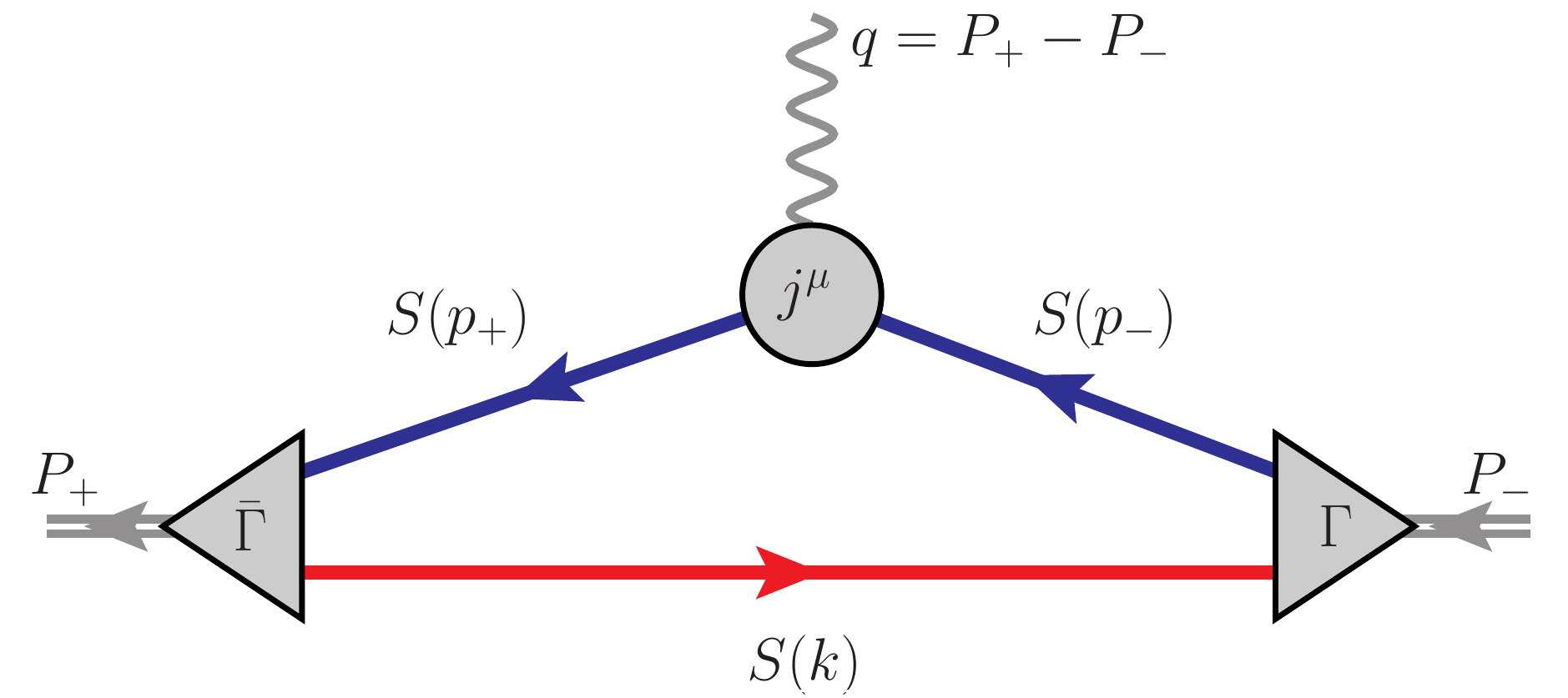}\vspace*{.5cm}
     \includegraphics[width=0.45\textwidth]{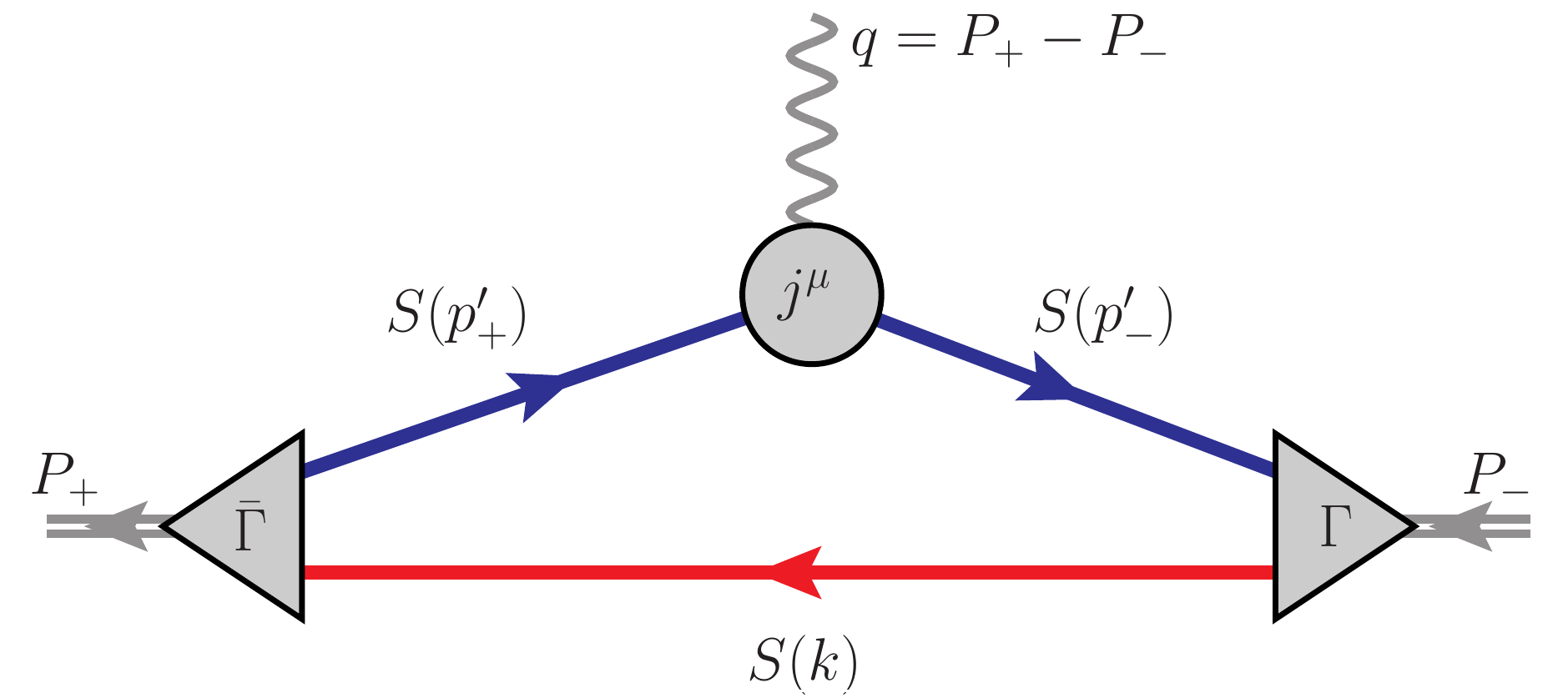}
\caption{(Color online) The two triangle diagrams for the electromagnetic pion form factor. The top diagram describes the interaction of the virtual photon with the $u$ quark, with the $\bar d$ quark  as a spectator; the bottom diagram represents the interaction of the virtual photon with the $\bar d$ quark, with the $u$ quark as the spectator. In both diagrams the $\bar d$ is represented by a $d$ quark traveling backward in time.}\label{fig:triangleDiagrams}
\end{center} 
\end{figure}

\section{\label{sec:triangleandingredients} The triangle diagram and ingredients}

\subsection{The triangle diagram in charge-conjugation-invariant CST}

The elastic electromagnetic form factor for a positively charged $\pi^+$ consisting of a $u$ and a $\bar d$ quark is obtained from the pion current, which---in impulse approximation---is the sum of two triangle diagrams, in which the photon couples either to the $u$ or the $\bar d$ quark. The two diagrams are depicted in Fig.~\ref{fig:triangleDiagrams}.

In the charge-conjugation invariant formulation of CST~\cite{Savkli:1999me,PhysRevD.89.016005} the $\pi^+$ current is given by
 \vspace{0.1in}
\begin{widetext}
\begin{eqnarray}\label{eq:Picurrent}
 J_{\pi^+}^\mu(P_+,P_-)=\mathrm e F_\pi (Q^2) (P_++P_-)^\mu
&=&\mathrm i\,\frac{2\,\mathrm e}{3}\int_{k0} \,
\mathrm {tr}\Big[\overline{\Gamma}(k,p_+) S(p_+) 
j^\mu (p_+,p_-) S(p_-) \Gamma(p_-, k)S(k)\Big]
\nonumber\\&&
-\mathrm i\,\frac{\mathrm e}{3}\int_{k0} \,
\mathrm {tr}\Big[\Gamma(k,p_-') S(p_-')
 j^\mu (p_-',p_+')S(p_+') \overline{\Gamma}(p_+',k)S(k)\Big]\,, 
\end{eqnarray}
 where $p_\pm=k+P_\pm$, $p'_\pm=k-P_\pm$, and $j^\mu (p_+,p_-)$ is the dressed current for off-shell quarks (defined below). We assume equal masses for the $u$- and $d$-quarks, so the $u$ and $d$ propagators are identical. Using this it has been shown in our earlier paper~\cite{PhysRevD.89.016006} that the second contribution to the form factor can be transformed into the first one, and the two can be added together. This gives
\begin{eqnarray}\label{eq:Picurrent2}
 J_{\pi^+}^\mu&&(P_+,P_-)
 =\mathrm i\,\mathrm e\int_{k0} \,
\mathrm {tr}\Big[\overline{\Gamma}(k,p_+) S(p_+) 
j^\mu (p_+,p_-) S(p_-) \Gamma(p_-, k)S(k)\Big]\,. 
\end{eqnarray}
\end{widetext}
 As in Eq.~(\ref{eq:CST-BSE}), the \lq\lq $k0$'' in Eq.~(\ref{eq:Picurrent}) stands for the charge-conjugation invariant CST prescription of how to perform the $k_0$-contour integration. It requires  taking all quark propagator-pole contributions of the $u$ and the $\bar d$ quark into account. The triangle diagram has six propagator poles in the complex $k_0$ plane, three  positive-energy poles in the lower- and three negative-energy poles in the upper-half plane. The \lq\lq $k0$'' prescription requires averaging over these poles. In the Breit frame, where
\bea
P_\pm&=&\left(P_0,{\bf 0},\pm\sfrac12 Q\right) 
\nonumber\\
q&=&\left(0,{\bf 0},Q\right) \label{eq:BreitFrame}
\eea 
with $P_0=\sqrt{\mu^2+\frac14Q^2}$,  $\mu$ the pion mass, and $Q$ the photon momentum transfer, the poles of the spectator $d$-quark are located at
$k_0=\eta E_k-\eta \,\mathrm i \epsilon$, with $\eta=\pm1$, $E_k = \sqrt{m^2+{\bf k}^2}$, and the poles of the active $u$-quark with momenta $p_-$ and $p_+$ are at 
\bea
k_0&=& \pm E_--P_0 \mp\mathrm i \epsilon \, ,
\nonumber\\
 k_0&=&\pm E_+-P_0\mp \mathrm i \epsilon \, ,
 \label{eq:k0pp}
 \eea
 respectively, where 
\bea
E_\pm=\sqrt{m^2+{\bf k}_\perp^2+\left(k_z\pm\frac Q2\right)^2}\, 
 \label{eq:E0pp}.
\eea
The active poles can be collectively written as $k_0=\eta E_{\eta'}-P_0-\eta \,\mathrm i \epsilon$, with $\eta=\pm1$ and $\eta'=\pm1$. Since the square roots in Eq.~(\ref{eq:E0pp}) are positive and $p_\pm=k+P_\pm$, $\eta=1$ denotes the positive-energy poles, and $\eta=-1$ denotes the negative-energy poles of the struck $u$-quark. The index $\eta'$ distinguishes between the poles of the active quark before and after its interaction with the photon.

\subsection{\label{subsec:pivertexfunction} Pion vertex function}

One input for the pion form factor calculation is the pion vertex function $\Gamma$. Instead of solving the full CST-BSE, which will be the subject of another paper, we use the approximated pion vertex function 
\begin{eqnarray}
  \Gamma(p_1,p_2)= G_0  h(p_1^2) h(p_2^2)\gamma^5 L(\rho^2) \, ,
 \label{eq:redvertex}
\end{eqnarray}
where $p_1=\rho+P/2$ and $p_2=\rho-P/2$ are the quark momenta in terms of the relative momentum $\rho$ and the total pion four-momentum $P=(P'_0,{\bf P}')$, with $P_0'=\sqrt{\mu^2+{\bf P}'^2}$; furthermore, $h(p_i^2)$ is a strong quark form factor normalized to $h(m_\chi^2)=1$, (where $m_\chi$ is the dressed quark mass in the chiral limit), and $G_0$ is the inverse norm of the wave function. The additional structure function $L(\rho^2)$ is a placeholder for the dynamically calculated pion vertex function, to be eventually obtained from the solution of the CST wave equation for the pion bound state.  All we currently know is that $L(\rho^2)=1$ in the chiral limit, and with this choice the form~(\ref{eq:redvertex}) coincides with the one introduced in Ref.~\cite{PhysRevD.89.016006}  and can be understood as a finite-pion-momentum extension of the chiral-limit pion vertex function in the pion rest frame 
\bea
\Gamma(\rho,\rho)=G_{\chi 0} h^2(\rho^2) \gamma^5 \, , \label{eq:vertex0}
\eea
which is obtained from solving the pion CST-BSE in the chiral-limit with a delta-function kernel of the form 
\bea
{\cal V}(\rho,\kappa)&=&\frac{C}{2m} H(p_1,p_2;\hat k,k_2) (2\pi)^3E_k \delta^3(\rho-\kappa)  \left(\gamma^\mu\otimes\gamma_\mu\right) 
\nonumber\\
&\to&\frac12 h^2(\rho^2)(2\pi)^3E_k\delta^3(\rho-\kappa)
 \left(\gamma^\mu\otimes\gamma_\mu\right) \, ,\label{eq:kernel} 
\eea
with $C$ the strength of this interaction ($C\rightarrow$ $m_\chi$ in the chiral limit \cite{PhysRevD.89.016005}), and with $H$ a shorthand notation for a product of four form factors
\bea
H(p_1,p_2;\hat{k}_1,k_2)&=&h(p_1^2)h(p_2^2)h(m^2)h(k_2^2)
\nonumber\\
&\to& h^2(\rho^2)\, . \label{eq:H}
\eea
The first line of Eq.~(\ref{eq:H}) gives the result for $H$ in the case when particle 1 is on-shell in the initial state [so that  $\hat k_1=(E_k,{\bf k}), k_1^2=m^2, \hat k_2=(E_k-P_0',{\bf k}-{\bf P}')$] and when both particles are off-shell in the final state. The second lines of the two Eqs.~(\ref{eq:kernel}) and (\ref{eq:H}) give the results in the chiral limit in the rest frame, when $m_0\to0, m\to m_\chi, \mu\to0$.  Note that a linear confining part of the interaction  has been omitted from (\ref{eq:kernel}); it was shown in Ref.~\cite{PhysRevD.89.016005} that this part does not contribute to the pseudoscalar bound-state equation in the chiral limit.

From here on, all quantities are assumed to be given in the chiral limit, except for the pion mass which is kept finite.

\subsection{\label{subsec:quarkcurrent} Quark current}
In a consistent pion form factor calculation, the dressed quark current should also be calculated from solving a (inhomogeneous) four-channel CST-BSE. This will, however, be the subject of a future paper and we use, for simplicity, the Ans\"atze proposed in Ref.~\cite{PhysRevD.89.016006}, which applies the framework introduced by Riska and Gross~\cite{Gro87} to ensure gauge invariance when strong form factors are attached to interaction vertices, such as in Eqs.~(\ref{eq:kernel}) and~(\ref{eq:H}). These form factors can be equivalently moved to the quark propagators, leading to so-called damped propagators
 \begin{eqnarray}
 \tilde S(p)=h^2(p^2) S(p)\,,\label{eq:Stilde}
\end{eqnarray}
reduced vertex functions
\begin{eqnarray}
 \Gamma_{\rm R}(p_1,p_2)=h^{-1}(p_1^2)\Gamma(p_1,p_2)h^{-1}(p_2^2)=G_0 \gamma_5\,,\label{eq:GammaR}
\end{eqnarray}
and reduced currents
\begin{eqnarray}
 j^\mu_{\rm R}(p',p)=h^{-1}(p'^2) j^\mu(p',p)h^{-1}(p^2)\,.\label{eq:jred}
\end{eqnarray}
The reduced current is then required to satisfy the Ward-Takahashi identity involving the damped propagators,
\begin{eqnarray}
 q^\mu j_{{\rm R}\mu } (p',p)=\widetilde S^{-1}(p)-\widetilde S^{-1}(p')\,,\label{eq:WTI}
\end{eqnarray}
to preserve gauge invariance.
The simplest possible  solution to this equation for pointlike quarks (i.e. quark form factors equal to 1 and no magnetic moment) is given by \cite{Gro96,PhysRevD.89.016006} 
 \begin{eqnarray}
 j_{\rm R}^\mu (p',p)
&=&f(p',p)
 \gamma^\mu
 + \delta(p',p)\Lambda(-p')\,\gamma^\mu\nonumber\\&&+\delta(p,p')\gamma^\mu\,\Lambda(-p)
  + g(p',p)\Lambda(-p')\,\gamma^\mu\,\Lambda(-p)\,.\nonumber\\\label{eq:jRcurrent}
\end{eqnarray}
The functions $f$, $\delta$, and $g$, are determined in terms of the strong form factor and the mass function through the Ward Takahashi identity: 
\begin{eqnarray}
  g(p',p)&=&\frac{4MM'}{h^2h'^2}\frac{(h^2-h'^2)}{(p'^2-p^2)}\,, \\
  \delta(p',p)&=&\frac{2M'}{h'^2}\frac{(M'-M)}{(p'^2-p^2)} \,,\\
  f(p',p)&=&\frac{M^2-p^2}{h^2(p'^2-p^2)}-\frac{M'^2-p'^2}{h'^2(p'^2-p^2)}\,,
  \end{eqnarray}
  where we have introduced the short-hand notation $h=h(p^2)$, $h'=h(p'^2)$, $M=M(p^2)$, and $M'=M(p'^2)$.
The mass function $M(p^2)$, calculated from the chiral limit CST-DE using the kernel in Eq.~(\ref{eq:kernel}), is equal to $m_\chi\,  h^2(p^2)$, which reduces the off-shell form factors $f$, $\delta$, and $g$ to
  
\begin{eqnarray}
  g(p',p)&\to&-2\delta(p',p)\to 4m_\chi^2\frac{(h^2-h'^2)}{(p'^2-p^2)}\,, \\
    f(p',p)&\to&\frac14g(p',p)+\frac{p'^2h^2-p^2h'^2}{h^2h'^2(p'^2-p^2)}\,.
   \label{eq:structurecurrent}
  \end{eqnarray}
  In order to study the influence of the running dressed quark mass $M(p^2)$ on the pion form factor, we will also consider in this paper the case of fixed dressed quark masses by setting $M(p^2)=m_\chi$. In this case the off-shell form factors reduce to
  
\begin{eqnarray}
  g(p',p)&\to&\frac{4m_\chi^2}{h^2h'^2}\frac{(h^2-h'^2)}{(p'^2-p^2)}\,, \\
  \delta(p',p)&\to&0\,,\\
  f(p',p)&\to&\frac{m_\chi^2-p^2}{h^2(p'^2-p^2)}-\frac{m_\chi^2-p'^2}{h'^2(p'^2-p^2)}\,.
  \end{eqnarray}
  
  \section{Contributions from the quark current} \label{sec:quark current}
  
Substituting~(\ref{eq:Stilde}),~(\ref{eq:GammaR}), and~(\ref{eq:jred}) into the pion current~(\ref{eq:Picurrent2}) gives 
\begin{eqnarray}
 J^\mu_{\pi^+}(P_+,P_-)&=&\mathrm i\,\mathrm e\int_{k0} G_0^2\frac{h^2h_+^2h_-^2 L_+L_- }{D\, D_+D_-}\nonumber\\
&&\times{\rm tr}\Big[\gamma^5  N(p_+)j^\mu_{\rm R}(p_+,p_-) N(p_-)\gamma^5\Lambda(k)\Big]\nonumber\\
&\equiv& J^\mu_{\pi} \, ,\label{eq:31c}
\end{eqnarray}
where we used the short-hand notation  $D=M^2-k^2$ and $D_\pm=M_\pm^2-p_\pm^2$ for the denominators of the propagators [with $M=M(k^2)$ and $M_\pm=M(p_\pm^2)$], and $h=h(k^2)$ and $h_\pm=h(p_\pm^2)$ for the strong form factors of spectator and active quarks, respectively, and  $L_\pm=L(\rho_\pm^2)$ (with $\rho_\pm=k+\frac12 P_\pm$); these additional structure functions are assumed to be functions of the relative momentum, but they might also depend on $p_\pm$ individually. 

 Next we insert~(\ref{eq:jRcurrent}) for the the quark current into (\ref{eq:31c}), which gives three contributions to the pion current associated with the off-shell form factors $f$, $\delta$, and $g$: 
\begin{eqnarray}
 J^\mu_{\pi}&=&J ^{f,\mu}_{\pi}+J^{\delta,\mu}_{\pi}+J ^{g,\mu}_{\pi}\,.
 \label{eq:decomposition}
\end{eqnarray}

In the following we analyze these three contributions separately and calculate for each of them the contributions from the spectator and the active poles.

Note that, from the  off-shell structure of the $g(p',p)$ term in Eq.~(\ref{eq:jRcurrent}), we can already anticipate that only the spectator pole will contribute in this case. This illustrates one of the advantages of breaking the calculation of the form factor into these three terms, instead of only concentrating on the total contribution of each of the six poles. The decomposition given by Eq.~(\ref{eq:decomposition}) turned out to be very useful for cross-checking our analytical and numerical calculations.

\subsection{$f$-term contribution}
The $f$ contribution to the pion current is
\begin{eqnarray}
 J^{f,\mu}_{\pi}
 &=&\mathrm i\,\mathrm e\int_{k0} G_0^2\frac{h^2 h^2_+h^2_-L_+L_- f(p_+,p_-)}{D\, D_+D_-}\nonumber\\
&&\times{\rm tr}\Big[N(p_+)\gamma^\mu N(p_-)\Lambda(-k)\Big]\,\nonumber\\
     &=&\mathrm i\,\mathrm e\int_{k0} G_0^2 \frac{ h^2 L_+L_-\Delta(k_+,k_-)\mathcal N^f}{(p_+^2-p_-^2)D}\,,
  \end{eqnarray}
where 
\begin{eqnarray}
     &&\Delta (p_+,p_-)= \frac{h_+^2}{M_+^2-p_+^2}-\frac{h_-^2}{M_-^2-p_-^2}\,,      \end{eqnarray}    
     and the trace is 
   \vspace{0.1in}
 \begin{widetext}
 
     \begin{eqnarray}
     \mathcal N^f&=&4 k^\mu (p_-\cdot p_+-M_- M_+)+4p_-^\mu (M M_+-k\cdot p_+)+4p_+^\mu (M M_--k\cdot p_-)\nonumber\\&=&
     4k^\mu (M M_+-M_- M_++M M_--k\cdot p_-+p_-\cdot p_+-k\cdot p_+)+4P_0^\mu (M M_++M M_--k\cdot p_--k\cdot p_+)\nonumber\\&&+2q^{\mu}(M M_--M M_++k\cdot p_+-k\cdot p_-)\,.
     \end{eqnarray}
  
  Note that $p_\pm^2=(k_0+P_0)^2-({\bf k}\pm\frac12 {\bf q})^2$, so that $p_+^2-p_-^2=-2k_zQ$, and excluding the $\mathcal N^f$ factor the integrand is even in ${\bf k}$. The $q^\mu$ term in $\mathcal N^f$ is odd in $k_z$ and hence this term integrates to zero -- a consequence of current conservation. For the $k^\mu$ term in $\mathcal N^f$ we introduce $2P_0^\mu\equiv (P_++P_-)^\mu$ and use the relation $k^\mu\to P_0^\mu k_0 /P_0$, which holds since the rest of the integrand is even in ${\bf k}$. Then, the remaining terms reduce to
  \begin{eqnarray}
  \mathcal N^f&=&
     4P_0^\mu \frac{k_0}{P_0} (M M_+-M_- M_++M M_--k^2+P_+\cdot P_-)+4P_0^\mu (M M_++M M_--2k^2-2k\cdot P_0)\nonumber\\
     &=&
     4P_0^\mu  \left[(M M_++M M_-)\left(1+\frac {k_0}{P_0}\right)-2k^2\left(1+\frac {k_0}{2P_0}\right)+(P_+\cdot P_--M_+ M_-)\frac{k_0}{P_0}-2k_0P_0\right]
     \nonumber\\
     &=&
     8P_0^\mu  \left\lbrace M\left[\frac{M_++M_-}{2}\right]-k^2+\left[M (M_++M_-)-M_+M_--k^2\right]\frac {k_0}{2P_0}-\frac{\mu^2k_0}{2P_0}\right\rbrace\,.
     \end{eqnarray} \end{widetext}  
     The $f$ contribution to the form factor then becomes
     \begin{eqnarray}
  F_{\pi}^f (Q^2) &=&-\mathrm i\,G_0^2
    \int_{k0} \frac{ h^2 L_+L_- \Delta(p_+,p_-)\chi(k_0,Q)}{k_zQP_0}\,,
  \label{eq:A23a}
  \end{eqnarray}
  where
    \begin{eqnarray} 
     &&\chi(k_0,Q)= \frac1D\left\lbrace P_0\left[M (M_++M_-)-2k^2\right]\right.\nonumber\\&&\left.\,\,\,+k_0\left[M (M_++M_-)-M_+M_--k^2-\mu^2\right]\right\rbrace\,.
    \end{eqnarray}
       For fixed dressed quark masses, $M=M_+=M_-\rightarrow m_\chi$, and $\chi(k_0,Q)$ reduces to
   \begin{eqnarray} 
     \chi(k_0,Q)\to 2P_0+k_0-\frac{\mu^2k_0}{m_\chi^2-k^2}\,.
    \end{eqnarray}  
Note that in the chiral limit {$\mu=0$}, such that the contribution of the spectator in the third term of this equation vanishes, and, in this limit, the entire result comes from the active quark poles in $\Delta$.
Next we perform the $k_0$ integration.

\subsubsection{Spectator pole contribution}
First we calculate the spectator pole contribution to $F_{\pi}^f (Q^2)$ in (\ref{eq:A23a}). Averaging over the poles at $k_0=\eta E_k$ gives
\begin{eqnarray}
  F_{\pi}^{f,s}(Q^2) &=&G_0^2
     \int\frac{\mathrm d^3 k}{(2\pi)^3 }\sum_{\eta,\eta'}\frac{ \eta'  \chi_{\eta}(Q) h^2_{\eta'\eta}L_{+,\eta}L_{-,\eta}}{2P_0k_zQ (M_{\eta'\eta}^2-p^2_{\eta'\eta})}\label{eq:A23e}\nonumber\\
  \end{eqnarray}  
where $p^2_{\eta'\eta}$ is the square of the active quark momentum, $p_{\eta'}^2$, evaluated at the spectator pole at $k_0=\eta E_k$, such that  
\bea
p^2_{\eta'\eta}&=&(\eta E_k+ P_0)^2-\Big({\bf k}+\eta'\frac12 {\bf q}\Big)^2\label{eq:petapeta}
\nonumber\\
&=&m_\chi^2+\mu^2+2\eta E_k P_0-\eta' k_z Q\,.
\eea
The relative momenta that appear as arguments of the $L$ functions become
\bea
\rho^2_{\eta'\eta}&=&\big(k+\frac12P_{\eta'}\big)^2=k^2+k\cdot P_{\eta'} +\frac{\mu^2}{4}
\nonumber\\
&=&m_\chi^2+\frac{\mu^2}{4}+\eta E_kP_0-\eta' k_z \frac{Q}{ 2}\, .
\eea
In (\ref{eq:A23e}) we have also used the abbreviations
\bea
&& h_{\eta'\eta}=h\big(p^2_{\eta'\eta}\big) \,,
\quad 
\nonumber\\
&& M_{\eta'\eta}=M\big(p^2_{\eta'\eta}\big)\,,
\nonumber\\
&& L_{\eta'\eta}=L(\rho^2_{\eta'\eta})
\,,\label{eq:A7}
\eea
 and 
\begin{eqnarray} 
  \chi_\eta(Q) &&\equiv  \chi(\eta E_k,Q) 
     \nonumber\\
     =&& \frac{P_0}{2E_k}\left[m_\chi (M_{+,\eta}+M_{-,\eta})-2m_\chi^2\right]
     \nonumber\\&&+\frac12\eta\left[m_\chi (M_{+,\eta}+M_{-,\eta})- M_{+,\eta}M_{-,\eta}-m_\chi^2-\mu^2\right] \,.\nonumber\\
    \end{eqnarray}    
 For fixed quark masses, $\chi_{\eta}(Q)\rightarrow -\eta \mu^2/2$, and the contribution to the form factor simplifies to
 \bea
F^{f,s}_{\pi}(Q^2)&\rightarrow&-G_0^2\int \frac{\mathrm d^3k}{(2\pi)^3}\frac{\mu^2}{4P_0k_zQ}
\nonumber\\&&\qquad\times
\sum_{\eta,\eta'}\frac{\eta\eta' h^2_{\eta'\eta}L_{+,\eta}L_{-,\eta}}{m_\chi^2-p^2_{\eta'\eta}}\,.
\eea
In Appendices \ref{sec:analysisspectatorfSQ} and~\ref{sec:analysisspectatorfLQ} 
we analyze, for fixed quark masses and $L(\rho^2)=1$, the small and large $Q^2$ behavior of the spectator contribution to the $f$ term. We find that the contribution has a $Q^{-2}$ falloff at large $Q^2$, consistent with the experimentally observed flattening of the $Q^2 F_{\pi}(Q^2)$ curve.

\subsubsection{Active pole contributions}
Next we calculate the active pole contributions by averaging over poles in the upper half plane at $k_0=-E_\pm-P_0$ and in the lower half plane at $k_0=E_\pm-P_0$. These cases are described by the subscripts $\{\eta'\eta\}$ in the general expression $k_0=\eta E_{\eta'}-P_0$. We need to evaluate, at the active particle pole $\{\eta'\eta\}$, the squares of the spectator momentum, $k$, of the active quark momentum, $p_{\eta''}$, and of the relative momentum, $\rho_{\eta''}$. They are denoted, respectively, by
\bea
\bar k^2_{\eta'\eta}&=&\big(\eta E_{\eta'} -P_0\big)^2-{\bf k}^2
\nonumber\\
&=&m_\chi^2+\mu^2-2\eta E_{\eta'}P_0 +\eta'k_zQ +\frac12 Q^2
\nonumber\\
(p^2_{\eta'})_{\eta'\eta}&=& m_\chi^2
\nonumber\\
(p^2_{-\eta'})_{\eta'\eta}&=&\big(\eta E_{\eta'})^2-k_\perp^2-\Big(k_z-\frac12\eta' Q\Big)^2
\nonumber\\
&=& m_\chi^2 +2\eta' k_zQ
\equiv \widetilde p^2_{\eta'} 
\nonumber\\
(\rho^2_{\eta'})_{\eta'\eta}&=& \Big(\eta E_{\eta'}-\frac12 P_0\Big)^2-k_\perp^2-\Big(k_z+\frac14\eta' Q\Big)^2
\nonumber\\&=&
\frac12\bar k^2_{\eta'\eta}+\frac12m_\chi^2-\frac14\mu^2
\nonumber\\
(\rho^2_{-\eta'})_{\eta'\eta}&=&\Big(\eta E_{\eta'}-\frac12 P_0\Big)^2-k_\perp^2-\Big(k_z-\frac14\eta' Q\Big)^2
\nonumber\\&=&
(\rho^2_{\eta'})_{\eta'\eta}+\eta' k_z Q 
\, . \label{eq:variables}
\eea
Since $(p^2_{\eta'})_{\eta'\eta}=m_\chi^2$, we need only a simplified notation for $(p^2_{-\eta'})_{\eta'\eta}$, namely
\bea
\bar M_{\eta'\eta}&=&M(\bar k_{\eta'\eta}^2)
\nonumber\\
\widetilde M_{\eta'}&=&M(\widetilde p_{\eta'}^2) \, .
\eea
$L_+L_-$ always occurs in the form of a symmetric product, such that
\bea
\bar L^2_{\eta'\eta}\equiv L\big[(\rho^2_{\eta'})_{\eta'\eta}\big] L\big[(\rho^2_{-\eta'})_{\eta'\eta}\big]\, .
\eea
At the pole indicated by $\{\eta'\eta\}$, $\chi(k_0,Q)$ is  
 \vspace{0.1in}
\begin{eqnarray} 
&&\bar \chi_{\eta'\eta}(Q)\equiv\chi(\eta E_{\eta'}-P_0,Q) 
\nonumber\\
&&=\frac{1}{\bar M^2_{\eta'\eta}-\bar k_{\eta'\eta}^2} \Big\{ P_0\left[\bar M_{\eta'\eta} (\widetilde M_{\eta'}+m_\chi)-2\bar k_{\eta'\eta}^2\right]
\nonumber\\
&&+(\eta E_{\eta'}-P_0)
[\bar M_{\eta'\eta} (\widetilde M_{\eta'}+m_\chi)- \widetilde M_{\eta'} m_\chi- \bar k_{\eta'\eta}^2-\mu^2]\Big\}.
\nonumber\\\label{eq:A24}
    \end{eqnarray} 
%
The $f$-term contribution of the active poles to the form factor then becomes
 \begin{equation}
  F_{\pi}^{f,a} (Q^2) =G_0^2
     \int\frac{\mathrm d^3 k}{(2\pi)^3 }\sum_{\eta,\eta'}\frac{ \eta' \bar \chi_{\eta'\eta}(Q)\bar h^2_{\eta'\eta}\bar L^2_{\eta'\eta}}{4k_zQ E_{\eta'}P_0} \, ,\label{eq:A23}
  \end{equation}    
where $\bar h_{\eta'\eta}=h(\bar k_{\eta'\eta}^2)$. In Appendices \ref{sec:fanalysisactiveSQ} and~\ref{sec:fanalysisactiveLQ} we analyze the small and large $Q^2$ behavior of this contribution for fixed dressed quark masses and $L(\rho^2)=1$. We obtained the somewhat unexpected result that the $f$-term contribution from the active quark poles has the same $Q^{-2}$ asymptotic falloff as the one from the spectator quark pole. This finding is independent of the precise functional form of the quark form factors and of details of the pion wave function.
 
 \subsection{$\delta$-term contribution}
 Next we look at the contributions from the $\delta$ terms in the quark current. Because the $\delta$ form factors are proportional to the difference of the mass functions, there are no contributions to this term in the case of fixed quark masses. For running quark masses the $\delta$-term contribution is given by
 \begin{eqnarray}
  J^{\delta,\mu}_{\pi}
  &=&\mathrm i \,\mathrm e\int_{k0}\frac{ G_0^2h_+^2h_-^2 h^2L_+L_-}{D D_+ D_-} \mathrm{tr} \Big\{N(p_+)
 \nonumber\\&&\times \Big[\delta(p_+,p_-)\Lambda(-p_+)\gamma^\mu
  +\delta(p_-,p_+)\gamma^\mu\Lambda(-p_-)\Big]
  \nonumber\\&&\times N(p_-)N(-k)\Big\}
  \,\nonumber\\
     &=&-\mathrm i \,\mathrm e\int_{k0} \frac{G_0^2h^2L_+L_-}{2k_zQD} \delta M\left\lbrace \frac{ h_-^2\mathcal N^\delta_-}{D_-} + \frac{h_+^2 \mathcal N^\delta_+}{D_+}\right\rbrace \qquad
  \end{eqnarray}
where $\delta M=M_+-M_-$, and  the traces are
  \begin{eqnarray}
     \mathcal N^\delta_-&=&{\rm tr}\Big\{\gamma^\mu(M_- + \slashed{p}_-)(M-\slashed{k})\Big\}
     \nonumber\\
     &=&
     4k^\mu ( M-M_-) +4 P_0^\mu M +2 q^\mu M\,,\nonumber\\
      \mathcal N^\delta_+&=&{\rm tr}\Big\{(M_++\slashed{p}_+)\gamma^\mu(M-\slashed{k})\Big\}
      \nonumber\\&=&
      4k^\mu ( M-M_+) +4 P_0^\mu M -2 q^\mu M\,.
    \end{eqnarray}
As before, the coefficient of the $q^\mu$ term is odd in $k_z$, and thus integrates to zero, a consequence of current conservation.   Since the factor multiplying $k^\mu$ is even in $k_z$,  we can substitute $k^\mu\rightarrow k_0P_0^\mu/P_0$ in the trace terms, which gives
    \begin{eqnarray}
     {\cal N}^\delta_+&\to&4P_0^\mu\left[ (M-M_+)\frac{k_0}{P_0} + M\right] \,,\nonumber\\
      {\cal N}^\delta_-&\to&4P_0^\mu\left[ (M-M_-)\frac{k_0}{P_0} + M\right]\,.
    \end{eqnarray}
   Then the contribution to the form factor becomes 
  \begin{eqnarray}
  F_{\pi}^\delta (Q^2)&=-&\mathrm i  \int_{k0} \frac{G_0^2 h^2L_+L_-}{k_zQP_0} \frac{\delta M}{M^2-k^2}\nonumber\\&&\times\left\lbrace \Sigma_1(p)(P_0+k_0)-\Sigma_2(p) k_0 \right\rbrace\,,\label{eq:AQFdFF}
  \end{eqnarray}
  where 
  \begin{eqnarray}
  \Sigma_1(p) &=& M\left[ \frac{ h_-^2}{M_-^2-p_-^2} + \frac{h_+^2 }{M_+^2-p_+^2}\right]\,,
  \nonumber\\
   \Sigma_2 (p) &=& \frac{ M_-h_-^2}{M_-^2-p_-^2} + \frac{M_+h_+^2 }{M_+^2-p_+^2}\,.
  \end{eqnarray}
It is explicit in the three last equations that both active and spectator quark poles contribute to the $\delta$ term of the pion form factor. Next we perform the $k_0$ integration.
   Averaging over the spectator poles at $k_0=\eta E_k$ gives the spectator contribution 
   \vspace{0.1in}
 \begin{widetext}
  \begin{eqnarray}
  F_{\pi}^{\delta,s} (Q^2)&=&
  G_0^2\int\frac{\mathrm d^3 k}{(2\pi)^3 } \frac{1}{4k_zQP_0E_k} \sum_{\eta,\eta'} h_{\eta'\eta}^2L_{+,\eta}L_{-,\eta}
  \left(\frac{ M_{ +,\eta}-M_{-,\eta}}{M_{\eta'\eta}^2-p_{\eta'\eta}^2}\right)\left[m_\chi P_0+\eta E_k(m_\chi-M_{\eta'\eta})\right]\,,\label{eq:AQFdFF}
  \end{eqnarray}
 and averaging over active quark poles at $k_0=\eta E_{\eta'}-P_0$ gives 
  \begin{eqnarray}
  F_{\pi}^{\delta,a} (Q^2)&=&-G_0^2\int \frac{\mathrm d^3 k}{(2\pi)^3}\frac{1} {4k_zQP_0}\sum_{\eta,\eta'} \frac{\eta'\bar h_{\eta'\eta}^2\bar L^2_{\eta'\eta}}{E_{\eta'}}
  \left(\frac{\widetilde M_{\eta'}-m_\chi}{\bar M_{\eta'\eta}^2-\bar k_{\eta'\eta}^2}\right)
  \left[ m_\chi P_0+\eta E_{\eta'}(\bar M_{\eta'\eta}- m_\chi)  \right]\,.
  \end{eqnarray}
 \end{widetext}

  \subsection{$g$-term contribution}

  The remaining contribution comes from the fully off-shell $g$ term in the quark current, which reads  
   \begin{eqnarray}
J^{g,\mu}_\pi&
=&\mathrm i\,\mathrm e\int_{k0} \frac{G_0^2 h^2 h_+^2 h_-^2L_+L_-} {D D_+ D_-}g(p_+,p_-)
\nonumber\\&&
\times\mathrm{tr} \Big\{N(p_+)
\Lambda(-p_+)\gamma^\mu\Lambda(-p_-) N(p_-) N(-k)\Big\}\nonumber\\
&=&-
\mathrm i\,\mathrm e\int_{k0} \frac{2G_0^2 h^2 (h_+^2-h_-^2)L_+L_- k^\mu}{k_z Q D }\,.
 \end{eqnarray}
  There is no contribution from the active quark poles to this term, because the denominators $D_+$ and $D_-$ cancel with $N(p_+)\Lambda(-p_+)$ and $\Lambda(-p_-)N(p_-)$, respectively. Hence, the only contribution comes from the spectator poles. Excluding the factor of $k^\mu$, the integrand is even in ${\bf k}$. Hence, we can replace $k^\mu\rightarrow k_0P_0^\mu/P_0$ and the $g$-term contribution to the pion form factor becomes 
  \begin{eqnarray}
F^{g}_\pi (Q^2)&=&-
\mathrm i\,G_0^2\int_{k0} \frac{ h^2 (h_+^2-h_-^2) L_+L_-k_0}{k_z Q D P_0 } \label{eq:gtermF}\,.
 \end{eqnarray}
 Then, averaging over the poles at $k_0=\eta E_k$ gives 
  \begin{equation}
F^{g}_\pi (Q^2)=G_0^2 \int\frac{\mathrm d^3 k}{(2\pi)^3 }\frac{ 1}{4k_z Q P_0 }
 \sum_{\eta,\eta'} \eta \eta' h_{\eta'\eta}^2 L_{+,\eta}L_{-,\eta}\,.
 \end{equation}
It is interesting to observe that the $g$-term contribution is identical for fixed and running quark masses, because the dependence on the mass function cancels out. In Appendices~\ref{sec:smallQgterm} and~\ref{sec:largeQgterm} we analyze, for $L(\rho^2)=1$, the small and large $Q^2$ behavior of this contribution. Similarly to what we found for the $f$-term, the $g$-term contribution also falls off as $Q^{-2}$, and is also independent of the functional form of the quark form factors.
 
\section{\label{sec:results} Results}
\begin{figure}
\begin{center}
    \includegraphics[width=0.45\textwidth]{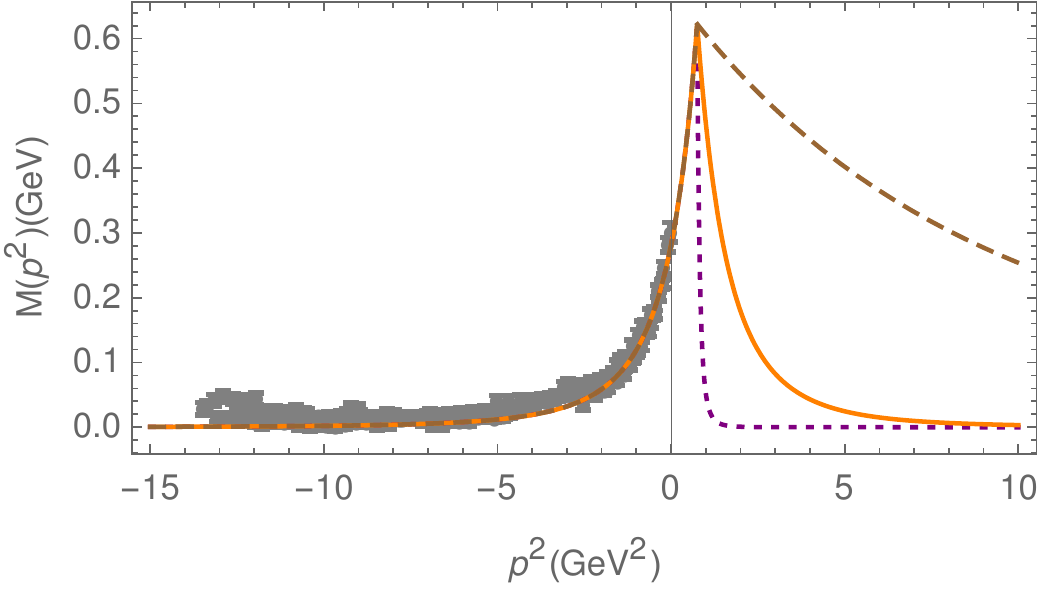}
\caption{(Color online) The chiral limit mass function $M(p^2)$, for $\alpha=3$ (dashed, brown), $\alpha=1$ (solid, orange), and  $\alpha=0.5$ (dotted, purple) compared with the lattice data~\cite{Bowman:2005vx} extrapolated to the chiral limit.}\label{fig:massfunction}
\end{center} 
\end{figure}

In our previous paper, Ref.~\cite{PhysRevD.89.016006}, on the pion form factor in the relativistic impulse approximation, we have used strong quark form factors of the simple form
\begin{eqnarray}
 h(p^2)=\left(\frac{\Lambda_\chi^2-m_\chi^2}{\Lambda_\chi^2-p^2}\right)^2\,. \label{eq:hffsimple}
\end{eqnarray}
Here $\Lambda_\chi$ is a cut-off parameter. Both, $m_\chi$ and $\Lambda_\chi$ are determined by a fit of the quark mass function $M(p^2)=m_\chi h^2(p^2)$ at negative $p^2$ to the lattice QCD data~\cite{Bowman:2005vx} extrapolated to the chiral limit~\cite{PhysRevD.89.016005}. The fit gives $\Lambda_\chi=2.04$ GeV and $m_\chi=0.308$ GeV. Note that $h(p^2)$ of (\ref{eq:hffsimple}) has a pole at $\Lambda_\chi^2=p^2$. Our previous calculation  only explored the values of $p_\pm^2$ at the RIA spectator particle pole (at $k_0=-E_k$), which, from Eq.~(\ref{eq:petapeta}), is bounded by
\bea
p_{\eta',-}^2&=&m_\chi^2+\mu^2 -2 E_k P_0-\eta' k_z Q
\nonumber\\
&\le& m_\chi^2+\mu^2<\Lambda_\chi^2\,.
\eea
 The contributions from the active particle poles  and the spectator pole at $k_0=E_k$ included in the present C-CIA calculation will probe the structure of the form factor $h$ also at large positive $p^2$ and will therefore depend on the definition of $h$ in this region.  To study this sensitivity, we adopt  a generalization of the piecewise form  proposed in~\cite{PhysRevD.89.016005}: 
\bea
 h(p^2)=\begin{cases}\displaystyle{\left(\frac{\Lambda^2_\chi-m_\chi^2}{\Lambda_\chi^2-p^2}\right)^2} & {\rm if}\; p^2<s_+\cr
\displaystyle{ \mathcal N(\alpha)\left(\frac{\alpha^2\Lambda_\chi^2-m_\chi^2}{\alpha^2\Lambda_\chi^2+p^2-2s_+}\right)^2} & {\rm if}\; p^2>s_+\,, \end{cases}  \label{eq:hffpiecewise}
 \eea
where $s_+<\Lambda_\chi^2$ is some fixed value (given below), and the normalization factor 
\bea
\mathcal N(\alpha)=\left[\frac{(\Lambda_\chi^2-m_\chi^2)(\alpha^2\Lambda^2-s_+)}{(\alpha^2\Lambda_\chi^2-m_\chi^2)(\Lambda^2-s_+)}\right]^2\,, \label{eq:Nalpha}
 \eea
makes $h(p^2)$ continuous at $p^2=s_+$. Note that this definition of $h(p^2)$ in the region $p^2>s_+$  ensures that it has no pole at $p^2=\Lambda_\chi^2$, or anywhere in the region if $\alpha^2\Lambda_\chi^2>s_+$.  Varying the value of the parameter $\alpha$ allows us to study the sensitivity of the pion form factor to the functional form of $h(p^2)$ in this region (where it is not constrained by lattice data). If $\alpha=1$, the new definition (\ref{eq:hffpiecewise}) is identical to the one proposed in \cite{PhysRevD.89.016005}, and we emphasize that 
when $p^2<s_+$, the form
(\ref{eq:hffpiecewise}) is identical to Eq.~(\ref{eq:hffsimple}) used in our  previous calculations of the pion form factor in the RIA~\cite{PhysRevD.89.016005} and, therefore, the results of these calculations remain unchanged by using~(\ref{eq:hffpiecewise}). We choose $s_+=(\Lambda_\chi-m_\chi)^2/4=0.752\, \mathrm{GeV}^2$. For $\alpha=1$, $h(p^2)$ is symmetric about $p^2=s_+$.
Figure~\ref{fig:massfunction} shows the mass function $M(p^2)=m_\chi h^2(p^2)$ for $\alpha=0.5,1,$ and 3 together with the lattice data extrapolated to the chiral limit.    

We have calculated the form factor for both, fixed dressed quark masses by setting $M(p^2)\rightarrow m_\chi$, and running dressed quark masses, with $M(p^2)=m_\chi h^2(p^2)$. In both cases we keep the strong quark form factors in the pion vertex functions and in the quark current as defined by Eq.~(\ref{eq:hffpiecewise}). All results presented here were obtained with $L(\rho^2)=1$.

\begin{figure}
\begin{center}
    \includegraphics[width=0.45\textwidth]{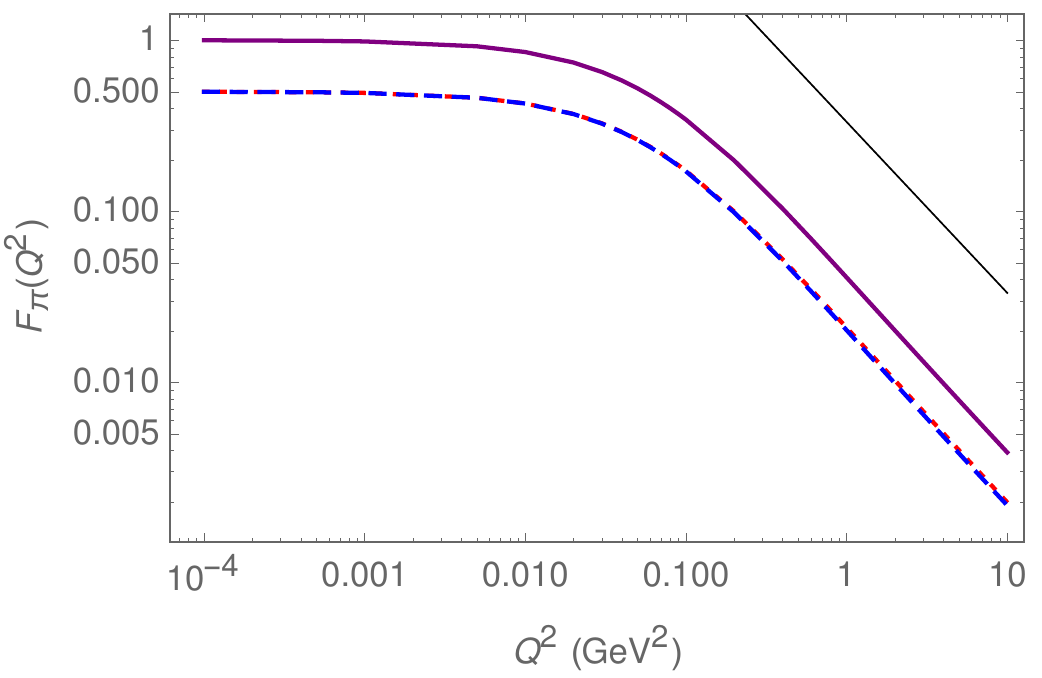}
    \includegraphics[width=0.45\textwidth]{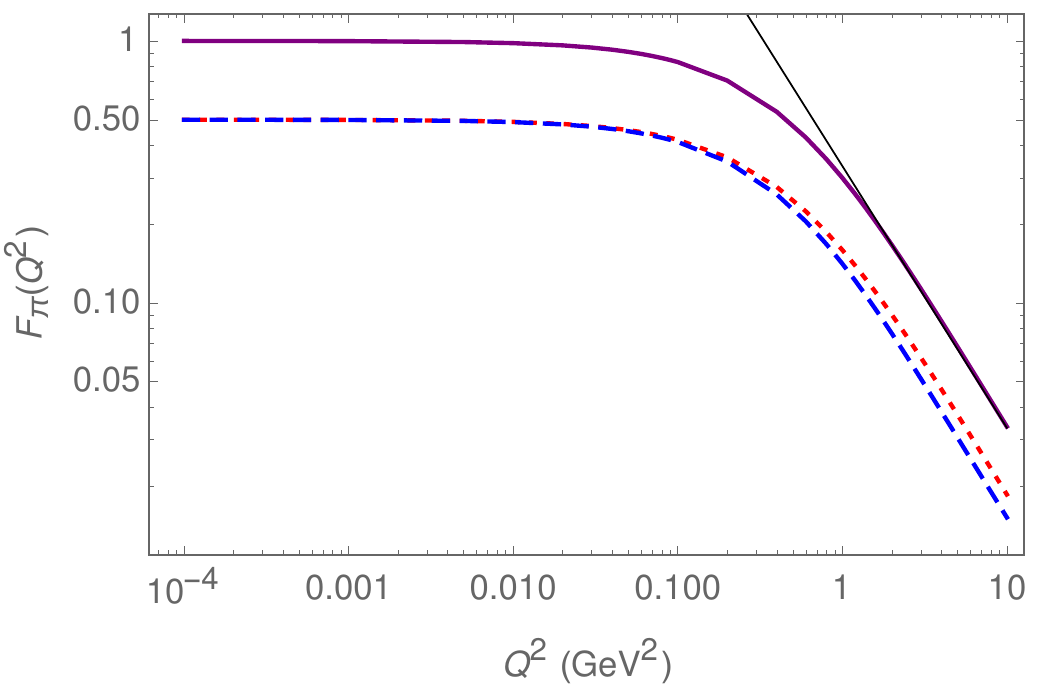}
    \includegraphics[width=0.45\textwidth]{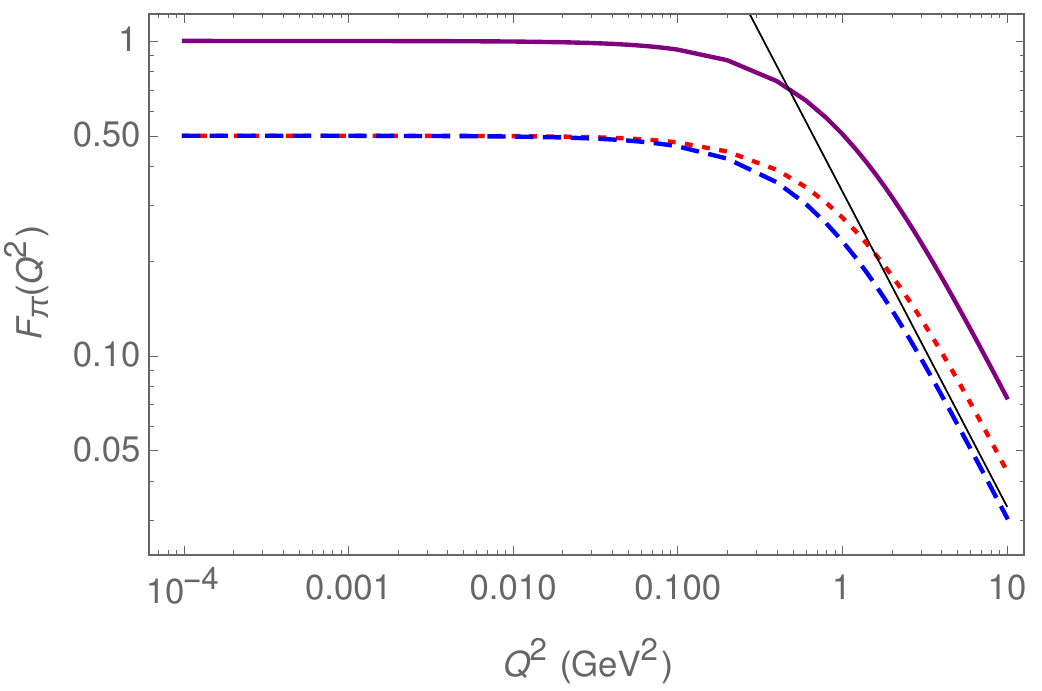}
\caption{(Color online) Comparison of the normalized spectator pole (red, dotted) and active pole (blue, dashed) contributions, calculated with
running quark masses, for $\mu=0.14$ GeV (top panel), for $\mu=0.42$ GeV (middle panel), and $\mu=0.60$ GeV (bottom panel). In each panel the sum of active and spectator pole contributions is the solid purple line which approaches 1 at small $Q^2$.  The asymptotic function 
$\frac1{3}Q^{-2}$, which fits the high $Q^2$ form factor data, is shown for comparison.}\label{fig:pifffixedm}
\end{center} 
\end{figure}

\begin{figure}
\begin{center}
\includegraphics[width=0.45\textwidth]{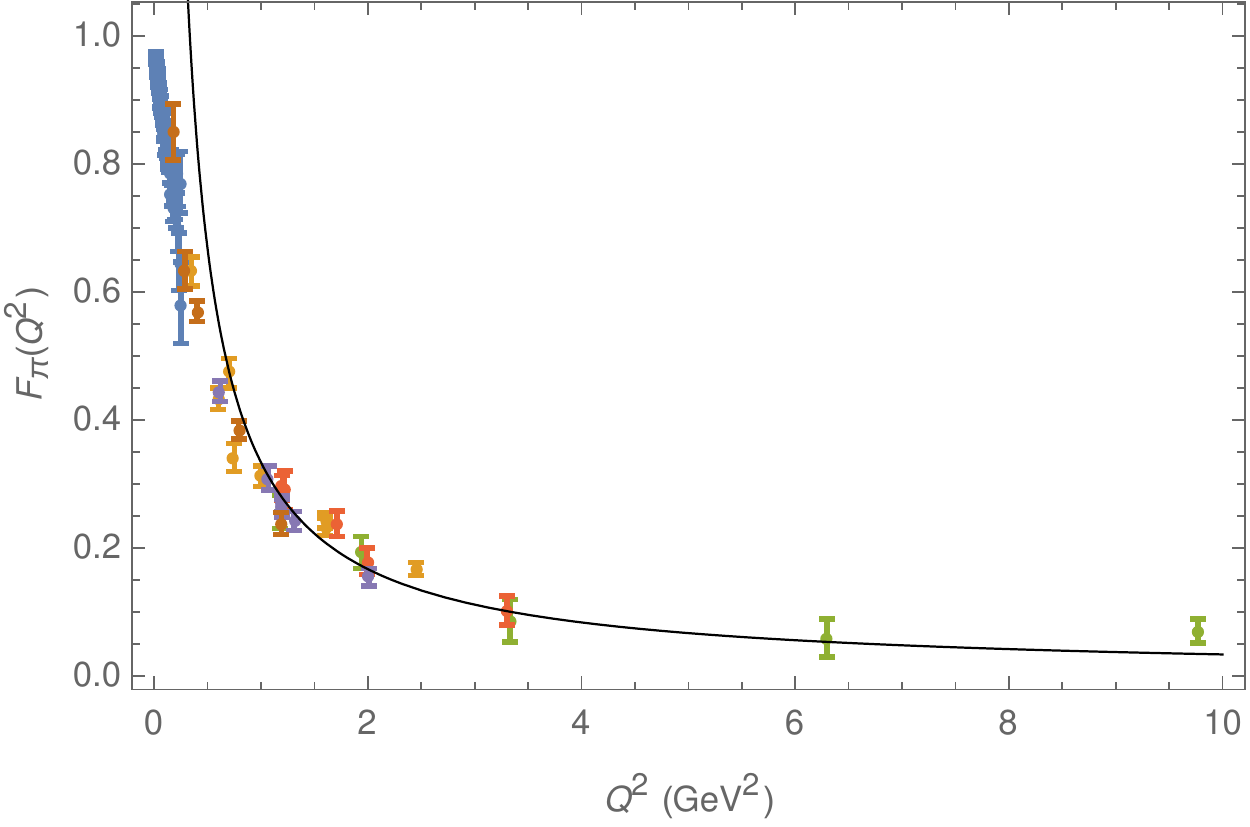} 
\caption{(Color online) The data~\cite{Amendolia:1986wj,Brown:1973wr,Bebek:1974iz,Bebek:1976ww,Bebek:1978pe,Volmer:2000ek,Horn:2006tm,Tadevosyan:2007yd,Huber:2008id} for the pion form factor shown against the simple model $F_\pi(Q^2)=\frac1{3}Q^{-2}$ (solid line).}\label{fig:data}
\end{center} 
\end{figure}

In Fig.~\ref{fig:pifffixedm} we compare the spectator and active quark pole contributions, calculated with running quark masses, for different values of the pion mass, namely $\mu=0.14$, $0.42$, and $0.6$ GeV. Also shown in this figure is the asymptotic function  $\frac13 Q^{-2}$ which gives a good fit to the pion form factor at large $Q^2$, as can be seen in Fig.~\ref{fig:data}. Comparing the slopes of this asymptotic form to the form factor curves  shows that all of the contributions have the correct $1/Q^2$ falloff at large $Q^2$. One notices in the results of Fig.~\ref{fig:pifffixedm} that, as the pion mass decreases, this asymptotic behavior sets in for smaller values of $Q^2$. Note also that both the spectator and active pole contribution are normalized by the same factor to ensure that $F_\pi^s(0)+F_\pi^a(0)=1$. 

\begin{figure}
\begin{center}
\includegraphics[width=0.45\textwidth]{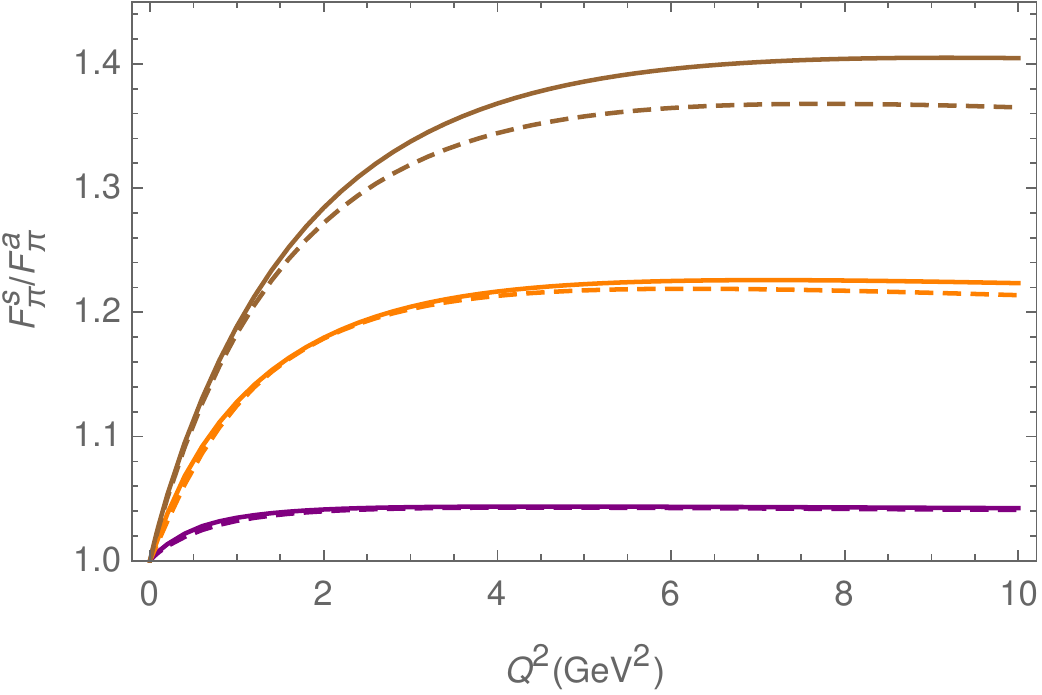}
\caption{(Color online) The ratio $F_\pi^s/F_\pi^a$ for fixed (dashed lines) and running (solid lines) quark masses, and different values of $\mu$. The pairs of curves, from top to bottom are the results obtained with $\mu=0.60$ (brown), $0.42$ (orange), and $0.14$ GeV (purple).}\label{fig:SpectOverAct}
\end{center} 
\end{figure}

Figure~\ref{fig:SpectOverAct} shows the ratio $F_\pi^s(Q^2)/F_\pi^a(Q^2)$ for fixed and running quark masses, and different values of $\mu$. It is a surprising result that these fixed and running mass calculations are very close to each other over the entire $Q^2$ range, with the possible exception for large $\mu$ and large $Q^2$, where the running mass results are larger by about 10\%.

\begin{figure}
\begin{center}
    \includegraphics[width=0.45\textwidth]{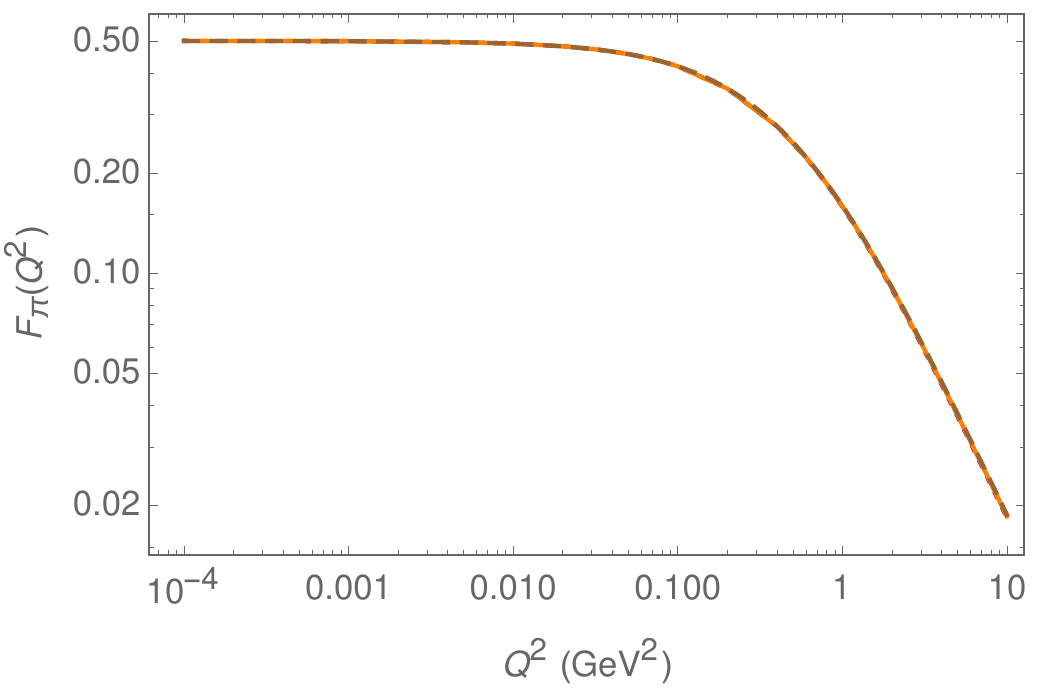}
    \includegraphics[width=0.45\textwidth]{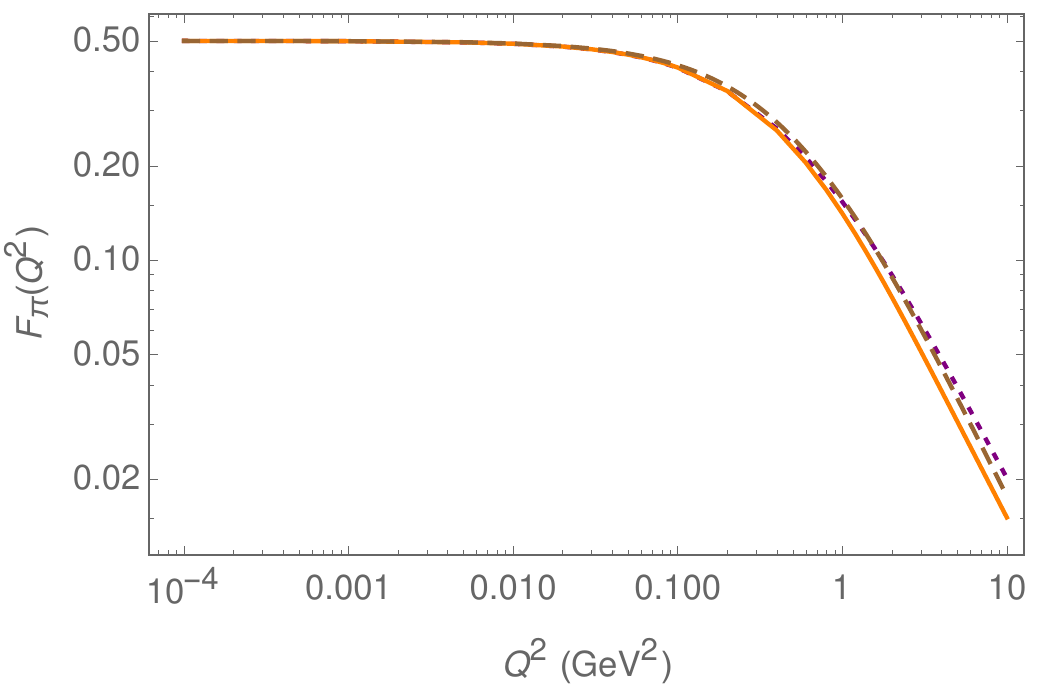}
\caption{(Color online) The pion form factor contributions calculated for $\alpha=3$ (dashed, brown), $\alpha=1$ (solid, orange), and  $\alpha=0.5$ (dotted, purple), with $\mu=0.42$ GeV and with running quark masses. The top panel shows the spectator and the bottom panel shows the active pole contributions.}\label{fig:alphavar}
\end{center} 
\end{figure}

We have also looked at the sensitivity of the pion form factor to the 
shape of the $h$ form factor in the timelike $p^2$ region, where it is not constrained by lattice data. In particular, we have varied the parameter $\alpha$ between 0.5 and 3. We find that the pion form factor is 
quite insensitive to these variations; only the active quark pole contributions at large $Q^2$ display any sensitivity at all. Figure~\ref{fig:alphavar} shows the results for $\mu=0.42$ GeV.

\section{\label{sec:summary} Summary and Conclusions}
The present study of the pion form factor using the Covariant Spectator Theory (CST) improves on our previous work~\cite{PhysRevD.89.016006}, which was done in the relativistic impulse approximation (RIA), where only one quark pole contribution to the triangle diagram for the pion current, the negative-energy spectator pole, was calculated. In the present work we take all six quark pole contributions into account. These are the positive and negative-energy spectator poles and the positive and negative-energy poles of the quark that interacts with the photon. The latter are referred to as the active poles. The inclusion of both, positive and negative-energy poles is necessary in order to preserve charge-conjugation invariance in CST~\cite{,PhysRevD.89.016005,Savkli:1999me}, hence we refer to the present calculation as the $C$-symmetric complete impulse approximation (C-CIA).

For the calculation of the pion form factor in C-CIA we use the same off-shell quark current as introduced in~\cite{PhysRevD.89.016006}. It satisfies the Ward-Takahashi identity and therefore our pion current is conserved. Our results for the pion form factor in C-CIA depend also on the pion mass $\mu$, similar to what was found in our previous RIA calculations~\cite{PhysRevD.89.016006}. As expected from the pole analysis therein, we find that for small $\mu$ the active quark contributions are as important as the spectator contributions, over the whole range of $Q^2$. For large $\mu$ and large $Q^2$, the active pole contributions are smaller than the spectator contributions by about 30\%.  For small $\mu$, the spectator and active pole contributions are nearly identical, not only in magnitude but also in shape, even for large $Q^2$. This is a surprising new result. Nevertheless, it confirms the correct large-$Q^2$ behavior obtained in~\cite{PhysRevD.89.016006}, which remains unchanged by the inclusion of the active poles, and sustains the flattening of $Q^2 F_{\pi}(Q^2)$ for large $Q^2$. 

We emphasize that the pion vertex function, as introduced in Ref.~\cite{PhysRevD.89.016006}, is a very simple Ansatz, that can be understood as a finite-pion-momentum extension of the pion vertex function in the chiral limit. The similarity between spectator and active pole contributions may be the result of using this simple pion vertex function with $L(\rho^2)=1$. 

We have also studied the effect of the running dressed quark masses on the pion form factor. We find that for small $\mu$ the results for the pion form factors calculated with running and with fixed dressed quark masses are nearly identical, and only for larger $\mu$ and $Q^2$ we observe some differences, which are, however, quite small.  

Related to this we have also investigated the sensitivity of the pion form factor on the strong quark form factors. In the C-CIA, the quark form factors, and accordingly the quark mass functions, are tested over the whole $p^2$ range. In particular, at positive $p^2$ where their shapes are not constrained by lattice QCD data, this offers the possibility to obtain important information about the mass function in this region. However, we find that for small pion mass the shape of the pion form factor, even in the small $Q^2$ region is quite insensitive to the functional form of the strong quark form factors. The early onset of the $Q^{-2}$ behavior for small $\mu$ may be connected to  this observed insensitivity. Only the active poles
for large $\mu$ and $Q^2$ show a small dependence. 

It should be mentioned that in this first C-CIA study of the pion form factor, we were rather constrained in our choices for the pion vertex function. In particular, at small pion masses our form factor persistently falls below the data at small $Q^2$. A more stringent comparison of our calculations with the data will be possible once precise pion vertex functions as solutions of the exact CST-BSE, as well as a dynamically calculated fully-dressed quark current, become available. Work on both subjects is currently under way.

\begin{acknowledgements}
E.P.B., M.T.P., and A.S., thank the Jefferson Lab Theory Center for its hospitality during recent visits when part of this work was performed. This work received financial support from Funda\c c\~ao para a Ci\^encia e a 
Tecnologia (FCT) under Grants No.~UID/FIS/00777/2013, No.~CERN/FP/123580/2011, and No.~SFRH/BPD/100578/2014. This work was also partially supported by  Jefferson Science
Associates, LLC, under U.S. Department of Energy Contract No. DE-AC05-06OR23177. The diagrams have been drawn with JaxoDraw \cite{Binosi200476}.
 \vspace{0.1in}
 
\end{acknowledgements}

\appendix

\section{Small and large $Q$ behavior of the pion form factor}

\subsection{\label{sec:analysgterm} $g$-term contribution}

Here we investigate the small and large $Q$ behavior of the $g$-term contribution $F^g_\pi (Q^2)$, Eq.~(\ref{eq:gtermF}).
\subsubsection{Small $Q$ behavior\label{sec:smallQgterm}}

It is easy to obtain the small $Q$ behavior if $\mu\ne0$ (the form factor diverges for all $Q$ when $\mu=0$). Then, to order $Q$,
\bea
h^2_{\eta'\eta}\simeq \hat h^2_{\eta}-\eta'k_zQ\frac{\mathrm d\hat h^2_{\eta}}{\mathrm d \hat p^2_\eta}\,,
\eea
where 
\bea
\hat p^2_\eta=\lim_{Q\to0} p_{\eta'\eta}^2=m_\chi^2+\mu^2+2\eta \mu E_k\,,\label{eq:A10}
\eea
 and
\bea
\hat h_{\eta}=h(\hat p^2_\eta)\,. \label{eq:A10b}
\eea
(Here symbols with ``hats'' should not be confused with on-shell four-vectors.) 
Note that $ \hat p^2_\eta\to\eta\infty$ as $k\to\infty$, {\it only\/} if $\mu\ne0$.  Since $\hat h_{\eta}$ does not depend on $\eta'$, the $\eta'$ sum contributes a factor of 2, and the small $Q$ limit becomes
\bea
\lim_{Q\to0}F_\pi^{g}(Q^2)&=&-G_0^2\int\frac{\mathrm d^3k}{(2\pi)^3} \frac{1}{2 \mu}\sum_{\eta} \eta \frac{\mathrm d\hat h^2_{\eta}}{\mathrm d \hat p^2_\eta} ,\qquad\quad \label{eq:A11}
\eea
where we have used that $P_0\to\mu$ as $Q\to0$.

\subsubsection{Large $Q$ behavior\label{sec:largeQgterm}}

At large $Q$ the arguments of $h$ can only be small (and hence the integrals large) if $k_z$ is very large (so that $k_z\sim E_k$). Expanding the arguments for large $k_z$ and $Q$ gives
 \bea
  p_{\eta'\eta}^2&\simeq&  m_\chi^2+\mu^2
+ Q\Big[\eta |k_z|+\eta\frac{m_\chi^2+k_\perp^2}{2|k_z|}-\eta' k_z\Big] 
\nonumber\\
&&+\frac{2\eta\mu^2 |k_z|}{Q}\equiv p^2_\infty\,.\qquad  \label{eq:A14}
 \eea
 For $p^2_\infty$ to be small at large $k_z$ requires that the sign of $k_z$ be chosen so that $\eta|k_z|-\eta' k_z=0$, which will be true only for positive $k_z$ if $\eta=\eta'$, and negative $k_z$ if $\eta=-\eta'$.  Hence the integrals for $\eta=\eta'$ run from $0 \le k_z< \infty$, and for $\eta=-\eta'$ from $-\infty<k_z\le 0$, and with this understanding, the arguments go to
 \bea
p^2_\infty&\to& \begin{cases} m_\chi^2+\mu^2+\eta Q[\frac{m_\chi^2+k_\perp^2}{2k_z} +\frac{2\mu^2 k_z}{Q^2}]\,,& \eta=\eta' 
\cr
m_\chi^2+\mu^2-\eta Q[\frac{m_\chi^2+k_\perp^2}{2k_z} +\frac{2\mu^2 k_z}{Q^2}]\,,  &\eta=-\eta'\,.
\end{cases}  
\nonumber\\&& \label{eq:A13}
 \eea    
We obtain 
\bea
\lim_{Q\to\infty}F^{g}_\pi(Q^2)=\frac{R^{g}}{Q^2}
\eea  
where, transforming $k_z=Qk''_z$, 
\bea
R^{g}&=&G_0^2\int\frac{\mathrm d^2k_\perp}{(2\pi)^3} \frac12\sum_{\eta}
\nonumber\\
&&\times\Bigg\{ \int_{0}^\infty \frac{\mathrm dk''_z}{k''_z} h^2[ p''^2_{\infty,\eta}]
-\int_{-\infty}^0 \frac{\mathrm dk''_z}{k''_z} h^2[ p''^2_{\infty,-\eta}]\Bigg\}
\nonumber\\
&=&G_0^2\int\frac{\mathrm d^2k_\perp}{(2\pi)^3} \int_0^\infty \frac{\mathrm dk''_z}{k''_z}
\sum_{\eta} h^2[p''^2_{\infty,\eta}],
\label{eq:A17}
\eea
where we introduced the convenient notation
\bea
 p''^2_{\infty,\eta}=m_\chi^2+2\eta\mu^2 \left(k''_z+\frac12\eta\right)+\eta\frac{m_\chi^2+k_\perp^2}{2k''_z} \, ,\qquad \label{eq:A16}
\eea
and, in the second equation, we changed the sign of $k''_z$, allowing the two terms to be combined into one (with a factor of 2).  
Note that the singularity that was at $k_z=0$ is now suppressed by the property $h(\pm\infty)=0$, and that, because $\mu\ne0$ the integrals converge at both $k''_z=0$ and $k''_z=\infty$.  The integrals peak at 
\bea
k''_{z}\Big|_{\rm peak}=\frac{\sqrt{m_\chi^2+k_\perp^2}}{2\mu}\, .
\eea
Notice that if $\eta=-1$, and $k''_z=1/2$, the argument becomes
\bea
p''^2_{\infty,-}\Big|_{\rm crit}=-k^2_\perp\, .
\eea

\subsection{\label{sec:fanalysis} $f$-term contribution}
Here we investigate, for fixed quark masses, the small and large $Q$ behavior of the $f$-term contribution $F^f_\pi(Q^2)$, Eq.~(\ref{eq:A23a}).
\subsubsection{\label{sec:fanalysisactiveSQ} Small $Q$ behavior of active pole contributions}

The small $Q$ behavior can be obtained by expanding $\bar h^2$ near $Q=0$ to order $Q$,
\be
\bar h^2_{\eta'\eta}\simeq \hat h^2_{-\eta}
+\frac{\eta' Q k_z}{E_k}(E_k-\eta\mu)\frac{\mathrm d \hat h^2_{-\eta}}{\mathrm d \hat p^2_{-\eta}} \,,
\ee
where $ \hat p^2_\eta$ and $\hat h_\eta$ were defined in Eq.~(\ref{eq:A10}). Using
\bea
\bar\chi_{\eta'\eta}(0)&=&\eta E_k+\mu-\frac{\mu(\eta E_k-\mu)}{2\eta E_k-\mu}
=\frac{2E_k^2}{2\eta E_k-\mu} \, , \qquad
\eea
which is nonsingular as long as $2m_\chi >\mu$, and taking the $Q\to0$ limit in the rest of (\ref{eq:A23}) gives
\bea
\lim_{Q\to0}F_{\pi}^{f,a}(Q^2)&=&G_0^2\int\frac{\mathrm d^3k}{(2\pi)^3}\frac1{\mu }
\nonumber\\&&\quad\times 
\sum_{\eta}\frac{\eta (\eta E_k-\mu)}{2\eta E_k-\mu}
\frac{\mathrm d \hat h^2_{-\eta}}{\mathrm d \hat p^2_{-\eta}}\qquad
\eea
where the two identical $\eta'$ terms have been combined, giving a factor of 2.  Changing $\eta\to-\eta$ in the sum gives a result which reduces to Eq.~(\ref{eq:A11}) if $\mu=0$,
\bea
\lim_{Q\to0}F_{\pi}^{f,a}(Q^2)&=&-G_0^2\int\frac{\mathrm d^3k}{(2\pi)^3}
\frac1{ \mu}
\nonumber\\&&\quad\times 
\sum_{\eta}\frac{\eta (\eta E_k+\mu)}{2\eta E_k+\mu}
\frac{\mathrm d \hat h^2_{\eta}}{\mathrm d\hat p^2_\eta}.  \qquad\quad \label{eq:A30}
\eea

 \subsubsection{\label{sec:fanalysisactiveLQ}Large $Q$ behavior of the active pole contributions}

The large $Q$ behavior of (\ref{eq:A23}) follows from arguments similar to those used for the study of $F^{g}_\pi$. Anticipating that the dominant contributions come from large $k_z$, we expand the arguments as follows:
\bea
\bar k^2_{\eta'\eta}
&\simeq&m_\chi^2+\mu^2-\eta Q |k_z+\frac12\eta'Q| +\eta'Q\Big[k_z +\frac12\eta'Q\Big]
\nonumber\\
&&-\eta Q\Bigg[\frac{m_\chi^2+k_\perp^2}{2|k_z+\frac12\eta' Q|} +\frac{2\mu^2|k_z+\frac12\eta' Q|}{Q^2}\Bigg]
\nonumber\\
& \equiv & \bar k^2_\infty\, .
\eea
Since $Q$ can be chosen to be positive, the argument is finite at large $Q$ only if
\bea
\eta  |k_z+\frac12\eta'Q| -\eta'\Big[k_z +\frac12\eta'Q\Big]=0 \, ,
\eea
which leads to the  conditions
\bea
k_z&>&-\frac12 \eta Q\qquad \eta=\eta'
 \eea 
 and
\bea 
k_z&<&\frac12\eta Q\qquad \eta=-\eta'\,.
\eea
Because of these conditions, when $\eta=\eta'$ it is convenient to shift the $k_z$ integration to $k_z\to k'_z-\frac12\eta Q$, so that $k_z+\frac12\eta Q\to k'_z$ and the region of integration is $k'_z>0$, while if $\eta=-\eta'$ the shift is $k_z\to k'_z+\frac12\eta Q$, so that $k_z-\frac12\eta Q\to k'_z$ and the region of integration is $k'_z<0$.  With these transformations, the arguments become
 \bea
\bar k^2_\infty&\to& \begin{cases} m_\chi^2+\mu^2-\eta Q[\frac{m_\chi^2+k_\perp^2}{2k'_z} +\frac{2\mu^2 k'_z}{Q^2}]\,,& \eta=\eta' 
\cr
m_\chi^2+\mu^2+\eta Q[\frac{m_\chi^2+k_\perp^2}{2k'_z} +\frac{2\mu^2 k'_z}{Q^2}]\,, & \eta=-\eta'\,.
\end{cases}
\nonumber\\&&
 \eea 
 Note that
 \bea
 \bar k^2_\infty=p^2_\infty(\eta\to-\eta,\eta'\to-\eta',k_z\to k_z')\, .
 \eea   
This means that the results obtained from the study of the $g$-term can be used here: the high $Q^2$ form factor is dominated by $k'_z>0$ when $\eta=\eta'$ and $k'_z<0$ when $\eta=-\eta'$.  In these regions, after shifting $k_z\to k'_z$ in the original definition (\ref{eq:A24}) (as discussed above), the large $Q$ behavior of $\bar \chi_{\eta'\eta}$ becomes 
 \bea
\bar \chi_{\eta'\eta}(Q)&\simeq& \eta |k'_z|+\frac12Q 
-\frac{2\mu^2Q|k'_z|(\eta |k'_z|-\frac12Q)}{Q^2\mathcal D(\eta)}
\nonumber\\&=&\frac1{Q^2\mathcal D(\eta)}\Big[ Q^2(m_\chi^2+k_\perp^2)(|k'_z|+\frac12\eta Q)\nonumber\\
&&\qquad +4\mu^2 k'^2_z(|k'_z|-\frac12\eta Q)\Big] 
\nonumber\\&=&\frac{Q}{\mathcal D(\eta)}\Big[ (m_\chi^2+k_\perp^2)(|k''_z|+\frac12\eta)\nonumber\\
&&\qquad +4\mu^2 k''^2_z(|k''_z|-\frac12\eta)\Big]
\nonumber\\
&\equiv&Q\bar \chi_R (\eta,|k''_z|)  \, ,
\eea 
where, in the second from last line, we anticipate the substitution $k'_z=Qk''_z$ to be made below, and  
\bea
Q^2\mathcal D(\eta)&=&\eta Q^2(m_\chi^2+k_\perp^2)+4\eta \mu^2 k'^2_z-2\mu^2Q|k'_z|\qquad
\nonumber\\
&=&Q^2\Big[\eta (m_\chi^2+k_\perp^2)+4\eta \mu^2 k''^2_z-2\mu^2|k''_z|\Big]\,.
\eea
Note that the reduced $ \bar \chi_R (\eta,|k''_z|)= \bar \chi_R (\eta,k''_z)$ if $k''_z>0$, and $ \bar \chi_R (\eta,|k''_z|)=\bar \chi_R (\eta,-k''_z)$ if $k''_z<0$.  Furthermore, it has the symmetry $\bar \chi_R (\eta,k''_z)= \bar \chi_R (-\eta,-k''_z)$.

Finally, at large $Q$ the result (\ref{eq:A23}) becomes
\bea
\lim_{Q\to\infty}F^{f,a}_{\pi}(Q^2)=\frac{R^{f,a}}{Q^2},
\eea  
where, making the substitutions discussed above,
\bea
R^{f,a}&=&G_0^2\int\frac{\mathrm d^2k_\perp}{(2\pi)^3} \sum_{\eta} \eta 
\nonumber\\
&&\times\Bigg\{ \int_{0}^\infty \frac{\mathrm dk''_z\bar \chi_R (\eta,k''_z)}{ 2 k''_z(k''_z-\frac12\eta)} h^2[p''^2_{\infty,-\eta}]
\nonumber\\&&
-\int_{-\infty}^0 \frac{\mathrm dk''_z  \bar \chi_R (\eta,-k''_z)}{ 2|k''_z|(k''_z+\frac12\eta)} h^2[ p''^2_{\infty,\eta}]\Bigg\}
\nonumber\\
&=&G_0^2\int\frac{\mathrm d^2k_\perp}{(2\pi)^3}\sum_{\eta}\eta \int_0^\infty \frac{\mathrm dk''_z  \bar \chi_R (\eta,k''_z)}{ k''_z(k''_z-\frac12\eta)}
 h^2[ p''^2_{\infty,-\eta}],
\nonumber\\
&& \label{eq:A17a}
\eea
where, in the second line, we have combined the two integrals over $k''_z$ by changing $k''_z\to -k''_z$ in the second integral.  When $\eta=1$, the integral has a principal value singularity at $k''_z=\frac12$, which can be easily integrated over.

\subsubsection{\label{sec:analysisspectatorfSQ}Small $Q$ behavior of the spectator pole contribution}

Evaluation of this contribution follows the discussion of the $g$-term contribution. We obtain immediately
\bea
\lim_{Q\to0}F_{\pi}^{f,s}(Q^2)&=&G_0^2\int\frac{\mathrm d^3k}{(2\pi)^3}  \frac{\mu}{2}\sum_{\eta}  \eta\frac{\mathrm d}{\mathrm d \hat p^2_\eta}\Big[\frac{\hat h^2_{\eta}}{m_\chi^2-\hat p^2_\eta}\Big] 
\nonumber\\
&=&G_0^2\int\frac{\mathrm d^3k}{(2\pi)^3} \frac1{2 \mu}\sum_{\eta} \frac{\eta}{2\eta E_k+\mu}
\nonumber\\
&&\qquad\times\Bigg\{\frac{\hat h^2_{\eta}}{(2\eta E_k+\mu)}-\mu
\frac{\mathrm d \hat h^2_{\eta}}{\mathrm d  \hat p^2_\eta}\Bigg\}.  \qquad\quad \label{eq:A36}
\eea
 The value of the total form factor at $Q=0$ is obtained by adding this to (\ref{eq:A11}) and (\ref{eq:A30}), giving
 \bea
 \lim_{Q\to0}&&F_{\pi}(Q^2)=G_0^2\int\frac{\mathrm d^3k}{(2\pi)^3} \frac{1}{2\mu}\sum_{\eta}  \eta
 \nonumber\\&&\qquad\times
 \Bigg\{\frac{\hat h^2_\eta}{(2\eta E_k+\mu)^2}-2\frac{\mathrm d\hat h^2_\eta}{\mathrm d \hat p^2_\eta}\Bigg\}\, .
 \eea
\vspace{-0.2in}
\subsubsection{\label{sec:analysisspectatorfLQ}Large $Q$ behavior of the spectator pole contribution}

The large $Q$ behavior also follows from the discussion of the $g$-term contribution. Using (\ref{eq:A13}) and repeating the steps leading to (\ref{eq:A17}) gives
\bea
\lim_{Q\to\infty}F^{f,s}_{\pi}(Q^2)=\frac{R^{f,s}}{Q^2}\,,\eea  
where, transforming $k_z=Qk''_z$, and using the definition (\ref{eq:A16}),
\bea
R^{f,s}&=&-G_0^2\int\frac{\mathrm d^2k_\perp}{(2\pi)^3} \frac{\mu^2}2\sum_{\eta}
\Bigg\{ \int_{0}^\infty \frac{\mathrm dk''_z}{k''_z} \frac{h^2[ p''^2_{\infty,\eta}]}{m_\chi^2- p''^2_{\infty,\eta}}
\nonumber\\&&
\qquad\qquad-\int_{-\infty}^0 \frac{\mathrm dk''_z}{k''_z} \frac{h^2[ p''^2_{\infty,-\eta}]}{m_\chi^2- p''^2_{\infty,-\eta}}\Bigg\}
\nonumber\\
&=&2G_0^2\int\frac{\mathrm d^2k_\perp}{(2\pi)^3} \int_0^\infty \mathrm dk''_z
\sum_{\eta}
\nonumber\\&&\qquad\times 
\frac{\eta\mu^2 h^2[p''^2_{\infty,\eta}]}{m_\chi^2+k_\perp^2+ 2\mu^2k''_z(2k''_z+\eta)}.
\label{eq:A17a}
\eea

\vspace*{0.1in} 

Combining the results for $R^{f,s}$, $R^{f,a}$, and $R^{g}$ gives the total coefficient for the $Q^{-2}$ falloff.

\vspace*{0.1in}

\bibliographystyle{h-physrev3}
\bibliography{PapersDB.bib}

\begin{thebibliography}{10}

\bibitem{Dudek:2012fk}
J.~Dudek {\em et~al.},
\newblock The European Physical Journal A {\bf 48} (2012).

\bibitem{Chang:2013qy}
L.~Chang, I.~C. Clo\"et, C.~D. Roberts, S.~M. Schmidt, and P.~C. Tandy,
\newblock Phys. Rev. Lett. {\bf 111}, 141802 (2013).

\bibitem{Godfrey}
S.~Godfrey and N.~Isgur,
\newblock Phys. Rev. D {\bf 32}, 189 (1985).

\bibitem{Eichten:1975}
E.~Eichten {\em et~al.},
\newblock Phys. Rev. Lett. {\bf 34}, 369 (1975).

\bibitem{Eichten:1978}
E.~Eichten, K.~Gottfried, T.~Kinoshita, K.~D. Lane, and T.~M. Yan,
\newblock Phys. Rev. D {\bf 17}, 3090 (1978).

\bibitem{Richardson:1978bt}
J.~L. Richardson,
\newblock Phys. Lett. B {\bf 82}, 272 (1979).

\bibitem{Edwards}
R.~G. Edwards, N.~Mathur, D.~G. Richards, and S.~J. Wallace,
\newblock Phys. Rev. D {\bf 87}, 054506 (2013).

\bibitem{Guo}
P.~Guo, J.~J. Dudek, R.~G. Edwards, and A.~P. Szczepaniak,
\newblock Phys. Rev. D {\bf 88}, 014501 (2013).

\bibitem{Brodsky:1997de}
S.~J. Brodsky, H.-C. Pauli, and S.~S. Pinsky,
\newblock Phys. Rept. {\bf 301}, 299 (1998).

\bibitem{Carbonell:1998rj}
J.~Carbonell, B.~Desplanques, V.~Karmanov, and J.~Mathiot,
\newblock Phys. Rept. {\bf 300}, 215 (1998).

\bibitem{Sales:1999ec}
J.~H.~O. Sales, T.~Frederico, B.~V. Carlson, and P.~U. Sauer,
\newblock Phys. Rev. C {\bf 61}, 044003 (2000).

\bibitem{Bars:1977ud}
I.~Bars and M.~B. Green,
\newblock Phys. Rev. D {\bf 17}, 537 (1978).

\bibitem{Amer:1983qa}
A.~Amer, A.~Le~Yaouanc, L.~Oliver, O.~Pene, and J.~C. Raynal,
\newblock Phys. Rev. Lett. {\bf 50}, 87 (1983).

\bibitem{LeYaouanc:1983it}
A.~Le~Yaouanc, L.~Oliver, O.~Pene, and J.~C. Raynal,
\newblock Phys. Lett. B {\bf 134}, 249 (1984).

\bibitem{Bicudo:1989sj}
P.~J. de~A.~Bicudo and J.~E. F.~T. Ribeiro,
\newblock Phys. Rev. D {\bf 42}, 1635 (1990).

\bibitem{Alkofer:2000wg}
R.~Alkofer and L.~von Smekal,
\newblock Phys. Rept. {\bf 353}, 281 (2001).

\bibitem{Maris:2003vk}
P.~Maris and C.~D. Roberts,
\newblock Int. J. Mod. Phys. E {\bf 12}, 297 (2003).

\bibitem{Fischer:2006ub}
C.~S. Fischer,
\newblock J. Phys. G {\bf 32}, R253 (2006).

\bibitem{Rojas:2013tza}
E.~Rojas, J.~de~Melo, B.~El-Bennich, O.~Oliveira, and T.~Frederico,
\newblock JHEP {\bf 1310}, 193 (2013).

\bibitem{PhysRevD.89.016006}
E.~P. Biernat, F.~Gross, M.~T. Pe\~na, and A.~Stadler,
\newblock Phys. Rev. D {\bf 89}, 016006 (2014).

\bibitem{PhysRevD.89.016005}
E.~P. Biernat, F.~Gross, M.~T. Pe\~na, and A.~Stadler,
\newblock Phys. Rev. D {\bf 89}, 016005 (2014).

\bibitem{Gross:1991te}
F.~Gross and J.~Milana,
\newblock Phys. Rev. D {\bf 43}, 2401 (1991).

\bibitem{Gross:1991pk}
F.~Gross and J.~Milana,
\newblock Phys. Rev. D {\bf 45}, 969 (1992).

\bibitem{Gross:1994he}
F.~Gross and J.~Milana,
\newblock Phys. Rev. D {\bf 50}, 3332 (1994).

\bibitem{Savkli:1999me}
C.~Savkli and F.~Gross,
\newblock Phys. Rev. C {\bf 63}, 035208 (2001).

\bibitem{Sal51}
E.~E. Salpeter and H.~A. Bethe,
\newblock Phys. Rev. {\bf 84}, 1232 (1951).

\bibitem{Gross:1993zj}
F.~Gross,
\newblock {\em Relativistic quantum mechanics and field theory}, revised ed.
  (New York, USA: Wiley-VCH, 1999).

\bibitem{Gro69}
F.~Gross,
\newblock Phys. Rev. {\bf 186}, 1448 (1969).

\bibitem{Gro87}
F.~Gross and D.~O. Riska,
\newblock Phys. Rev. C {\bf 36}, 1928 (1987).

\bibitem{Gro96}
Y.~Surya and F.~Gross,
\newblock Phys. Rev. C {\bf 53}, 2422 (1996).

\bibitem{Bowman:2005vx}
P.~O. Bowman {\em et~al.},
\newblock Phys. Rev. D {\bf 71}, 054507 (2005), hep-lat/0501019.

\bibitem{Amendolia:1986wj}
S.~Amendolia {\em et~al.},
\newblock Nucl. Phys. B {\bf 277}, 168  (1986).

\bibitem{Brown:1973wr}
C.~N. Brown {\em et~al.},
\newblock Phys. Rev. D {\bf 8}, 92 (1973).

\bibitem{Bebek:1974iz}
C.~J. Bebek {\em et~al.},
\newblock Phys. Rev. D {\bf 9}, 1229 (1974).

\bibitem{Bebek:1976ww}
C.~J. Bebek {\em et~al.},
\newblock Phys. Rev. D {\bf 13}, 25 (1976).

\bibitem{Bebek:1978pe}
C.~J. Bebek {\em et~al.},
\newblock Phys. Rev. D {\bf 17}, 1693 (1978).

\bibitem{Volmer:2000ek}
J.~Volmer {\em et~al.},
\newblock Phys. Rev. Lett. {\bf 86}, 1713 (2001).

\bibitem{Horn:2006tm}
T.~Horn {\em et~al.},
\newblock Phys. Rev. Lett. {\bf 97}, 192001 (2006).

\bibitem{Tadevosyan:2007yd}
V.~Tadevosyan {\em et~al.},
\newblock Phys. Rev. C {\bf 75}, 055205 (2007).

\bibitem{Huber:2008id}
The Jefferson Lab F{$_\pi${}} Collaboration, G.~M. Huber and E.~Al,
\newblock Phys. Rev. C {\bf 78}, 045203 (2008).

\bibitem{Binosi200476}
D.~Binosi and L.~Theussl,
\newblock Comp. Phys. Comm. {\bf 161}, 76  (2004).

\end{thebibliography}

\end{document}